\documentclass[11pt]{article}
\usepackage{amssymb}
\usepackage{amsmath}
\usepackage[all]{xy}
\input{epsf}
\usepackage{epsfig}
\usepackage{pifont}
\numberwithin{equation}{section}

\def\be{\begin{equation}}
\def\ee{\end{equation}}
\def\ba{\begin{align}}
\def\ea{\end{align}}
\def\beq{\begin{eqnarray}}
\def\eeq{\end{eqnarray}}

\def\p{\partial}

\def\T{\mathcal{T}}
\def\F{\triangleright}
\def\SF{\substack{\triangleleft \\ \triangleright}}

 \input{epsf}

 \usepackage{epsfig}

\begin{document}

\title{\Large{\bf 
First-principles derivation of the AdS/CFT Y-systems}} 
\author{Raphael Benichou}
\maketitle
\begin{center}
 Theoretische Natuurkunde, Vrije Universiteit Brussel and \\
The International Solvay Institutes,\\
 Pleinlaan 2, B-1050 Brussels, Belgium \\
 \textsl{raphael.benichou@vub.ac.be}
\end{center}

 \begin{abstract}
We provide a first-principles, perturbative derivation of the $AdS_5$/$CFT_4$ Y-system 
that has been proposed to solve the spectrum problem of $\mathcal{N}=4$ SYM.
The proof relies on the computation of quantum effects in the fusion of some loop operators, namely the transfer matrices. 
More precisely we show that the leading quantum corrections in the fusion of transfer matrices induce the correct shifts of the spectral parameter in the T-system.
As intermediate steps we study UV divergences in line operators up to first order and compute the fusion of line operators up to second order for the pure spinor string in $AdS_5\times S^5$.
We also argue that the derivation can be easily extended to other integrable models, some of which describe string theory on $AdS_4$, $AdS_3$ and $AdS_2$ spacetimes.
\end{abstract}

\newpage

\tableofcontents

\newpage


\section{Introduction}

The AdS/CFT correspondence \cite{Maldacena:1997re}\cite{Gubser:1998bc}\cite{Witten:1998qj} implies that type IIB string theory in $AdS_5 \times S^5$ is equivalent to $\mathcal{N}=4$ Super-Yang-Mills in four dimensions.
In the classical string theory limit, or equivalently in the planar gauge theory limit, integrable structures appear.
This has lead to impressive progress in the understanding of this system (see \cite{Beisert:2010jr} for a review).

The AdS/CFT dictionary relates the energy of string states in the bulk to the conformal dimensions of operators on the boundary.
A set of equations known as the Y-system has been put forward in \cite{Gromov:2009tv} to solve the spectrum problem of planar $\mathcal{N}=4$ SYM, or equivalently of string theory in $AdS_5 \times S^5$.
The goal of the present article is to make a new step towards a definite proof of the validity of this set of equations.

There is by now solid evidence in favor of the validity of the Y-system.
It reproduces the results of the Asymptotic Bethe Ansatz \cite{Beisert:2006ez}, but it does not suffer from the same limitations.
For instance it contains \cite{Gromov:2009tq}\cite{Gromov:2010vb} the spectrum of the quasi-classical string at large 't Hooft coupling (see e.g. \cite{Gromov:2007ky}).
Even more impressively, it lead to correct predictions for the dimension of the Konishi operator both at large \cite{Gromov:2009zb} and at small \cite{Arutyunov:2010gb} 't Hooft coupling.

In order to claim that the spectrum problem for $\mathcal{N}=4$ SYM has been definitively solved, it would be comfortable to have a proof of the validity of the Y-system.
At that point the only known derivation of the Y-system relies on the Thermodynamic Bethe Ansatz \cite{Zamolodchikov:1989cf} (see e.g. \cite{Bajnok:2010ke} for a review). This approach was studied in \cite{Gromov:2009bc}\cite{Bombardelli:2009ns}\cite{Arutyunov:2009ur}.
The Thermodynamic Bethe Ansatz has been very successful and lead to numerous remarkable results.
However this method relies on several crucial assumptions. 
In the first place, one has to assume quantum integrability of the theory.
Then one needs the ``string hypothesis'': the spectrum of excitations that contribute to the thermodynamic limit of the theory essentially has to be guessed.
Most importantly, this method only gives the ground state energy. The spectrum of excited state can be obtained by analytic continuation, but the reason why this works is not understood.

In this paper, we will initiate a different approach to derive the Y-system from first principles.
We will use only elementary tools of two dimensional conformal field theory; in this aspect this article can be related to the seminal work of \cite{Bazhanov:1994ft} where the Y-system was derived for the minimal models.
We will be able to prove the validity of the Y-system up to first non-trivial order at large 't Hooft coupling.

\paragraph{The idea of the proof.}
Up to a change of variables, the Y-system can be rewritten as a T-system, also known as the Hirota equation:
\be\label{Tsystem} \mathcal{T}_{a,s}(u + 1) \mathcal{T}_{a,s}(u - 1) = 
  \mathcal{T}_{a+1,s}(u+1)\mathcal{T}_{a-1,s}(u-1) +  \mathcal{T}_{a,s+1}(u-1)\mathcal{T}_{a,s-1}(u+1) \ee
In the above equation, $u$ is a spectral parameter and the indices $(a,s)$ are integers that label representations of the global symmetry group of the system. In the case at hand this group is $PSU(2,2|4)$. These labels take values in a T-shaped lattice. More details are given in section \ref{proofTsys}.

The T-functions are expected to be related to special line operators of the string worldsheet theory, namely the transfer matrices (see e.g. \cite{Gromov:2010vb}).
The transfer matrices play a central role in the study of classical integrability. Indeed these operators code an infinite number of conserved charges.
By definition, the transfer matrix \eqref{defTransfer} is the supertrace of the monodromy matrix. 
The monodromy matrix itself is an element of the supergroup $PSU(2,2|4)$. 
Thus the classical transfer matrix is a supercharacter.
If the shifts of the spectral parameter are neglected, the T-system \eqref{Tsystem} reduces to a character identity that is known to hold (see e.g.  \cite{Gromov:2010vb}\cite{Kazakov:2007na}).
We deduce that the shifts of the spectral parameter come from some kind of quantum effects.

In this paper we take the identification the T-functions to the transfer matrices seriously.
The T-system is promoted to an operator identity between quantum transfer matrices.
We postulate that the product appearing in the T-system is the fusion of line operators.
The process of fusion of line operators involve quantum effects, that are responsible for the appearance of the shifts in the T-system.
We will show that this picture is indeed correct up to first non-trivial order in the large 't Hooft coupling expansion.
More precisely, we will compute the leading quantum correction in the fusion of two transfer matrices, and show that it correctly gives the shifts of the T-system at first order.
The same strategy was previously successfully applied for the sigma-model on the supergroup $PSl(n|n)$ in \cite{Benichou:2010ts}.

\paragraph{Organization of the paper.}
In section \ref{worldsheetTheory} we describe the features of the pure spinor string on $AdS_5\times S^5$ that are relevant for our purposes. We also introduce the relevant line operators.
In section \ref{fusion} we present the central computation of this work: we study the fusion of line operators up to second order in perturbation theory.
In section \ref{proofTsys} we make good use of this computation to deduce the validity of the T-system up to first order in perturbation theory.
Section \ref{extensions} contains a discussion of the extension of this method to other integrable models.
Eventually final remarks are gathered in section \ref{conclusion}.

In order to keep the bulk of the paper as readable as possible, most of the details of the computations are gathered in the appendices. Appendix \ref{conventions} contains the conventions. In Appendix \ref{covCurrents} we revisit the computation of the current-current OPEs in the pure spinor formalism using a novel and efficient method. In Appendix \ref{computationDiv} we study the UV divergences in line operators. Eventually Appendix \ref{comFusion} contains the computations relevant for the fusion of line operators.


\section{The pure spinor string on $AdS_5 \times S^5$}\label{worldsheetTheory}

To describe superstring theory in $AdS_5 \times S^5$ we will use the pure spinor formalism.
This choice is a matter of convenience.
Indeed the computations of section \ref{fusion} are simpler in the conformal gauge, where target-space covariance is preserved.

In this section we introduce the pure spinor string on $AdS_5 \times S^5$. 
We only discuss the features of this formalism that are relevant for the purpose of this paper. 
A more detailed discussion can be found for instance in \cite{Berkovits:2000fe}\cite{Berkovits:2004xu}\cite{Mazzucato:2011jt}.
We introduce the flat connection, and show that the commutator of equal-time connections can be written in the canonical form of a $(r,s)$ system. 
We also introduce the line operators that are defined as the path-ordered exponential of the line integral of the flat connection. Finally we discuss the UV divergences that appear in these line operators because of quantum effects.
Most of the results discussed in this section have appeared before in the literature.
Some new results are presented concerning the current algebra, the commutator of equal-time connections and the renormalization of the line operators.


\subsection{Generalities}

The target-space $AdS_5\times S^5$ is embedded in a superspace with 32 supercharges. It is realized as the supercoset $PSU(2,2|4)/SO(4,1)\times SO(5)$. 
The Lie superalgebra $\mathcal{G}=psu(2,2|4)$ admits the action of a $\mathbb{Z}_4$ automorphism. This automorphism induces a $\mathbb{Z}_4$ grading on the elements of the Lie superalgebra. We can decompose the Lie algebra $\mathcal{G}$ according to this grading:
\be\label{Z4decompo} \mathcal{G} = \mathcal{H}_0 \oplus \mathcal{H}_1 \oplus \mathcal{H}_2 \oplus \mathcal{H}_3 \ee
where the subscript gives the $\mathbb{Z}_4$ grade. Bosonic (respectively fermionic) generators of the Lie superalgebra have an even (respectively odd) grade.

\paragraph{The action.}
Let us introduce the currents $J$ and $\bar J$ defined in terms of the group element $g \in PSU(2,2|4)$ as:
\be\label{J=gdg} J = g^{-1} \p g \quad ; \quad \bar J =  g^{-1} \bar \p g \ee
They take values in the Lie superalgebra $\mathcal{G}$.
We decompose the current $J$ according to the $\mathbb{Z}_4$ grading of the Lie superalgebra:
\be J = J_{0} + J_{1} + J_{2} + J_{3} \ee
and similarly for $\bar J$.
Let us also introduce the bosonic pure spinor ghosts $\lambda, \hat \lambda$ as well as their conjugate momenta $w, \hat w$. They expand on the fermionic generators of the superalgebra with the following gradings: $\lambda, \hat w \in \mathcal{H}_1$ and $\hat \lambda, w \in \mathcal{H}_3$. 
The ghosts satisfy the pure spinor constraint:
$ \lambda \gamma^\mu \lambda = 0 =  \hat \lambda \gamma^\mu \hat \lambda$,
where the $\gamma^\mu$'s are the $SO(9,1)$ gamma matrices.
The pure spinor Lorentz currents are:
\be N=-\{w,\lambda\} \qquad ; \qquad \hat N = -\{\hat w, \hat \lambda\} \ee
The action reads:
\begin{align}\label{action} S =& \frac{R^2}{4\pi } STr \int d^2 z \left( J_2 \bar J_2 + \frac{3}{2} J_3 \bar J_1 + \frac{1}{2} \bar J_3 J_1 \right)  \cr
&+  \frac{R^2}{2 \pi} STr \int d^2 z \left( N \bar J_0 + \hat N J_0 - N \hat N + w \bar \p \lambda + \hat w \p \hat \lambda \right)
\end{align}
The first line of the action contains a kinetic term both for the bosonic and fermionic target space coordinates. This implies in particular that the model does not exhibit kappa-symmetry, contrary to the Green-Schwarz string.
The radius of the target space is denoted by $R$ in units of the string length.
Later on we will work perturbatively in a large radius expansion: the small parameter is $R^{-2}$.

\paragraph{Gauge symmetry.}
The action \eqref{action} admits a $\mathcal{H}_{0}$ gauge symmetry:
\be \delta g =  g h_0, \quad \delta \lambda = [\lambda,h_0], \quad \delta \hat \lambda = [\hat \lambda,h_0],
 \quad \delta w = [w,h_0], \quad \delta \hat w = [\hat w,h_0] \ee
The holomorphic currents transform as:
\be i\neq 0:\ \delta J_{i} = [J_{i} ,h_0]\quad ;  \quad
\delta J_{0} = \p h_0 + [J_{0} ,h_0]\quad ; \quad
\delta N = [N,h_0]
\ee
and similarly for the anti-holomorphic currents.
We introduce the associated covariant derivative:
\be \nabla = \p + [J_{0},\cdot ] \qquad ; \qquad  \bar \nabla = \bar \p + [\bar J_{0},\cdot ] \ee

\paragraph{Parity.}
The model enjoys a $\mathbb{Z}_2$ symmetry that exchanges holomorphic and anti-holomorphic worldsheet coordinates. It also flips the grade of the fermionic elements of the Lie superalgebra: the subalgebras $\mathcal{H}_1$ and $\mathcal{H}_3$ are exchanged.

\paragraph{The Maurer-Cartan equation.}
A consequence of \eqref{J=gdg} is that the current satisfies the Maurer-Cartan equation:
\be\label{MC} \p \bar J - \bar \p J + [J, \bar J] = 0 \ee
This is a crucial equation that is essentially responsible for the integrable properties of the model.

\paragraph{Equations of motion.}
The equations of motion combined with the Maurer-Cartan equation lead to:
\be\label{MC+EOM}
\begin{array}{lll}
\bar \nabla J_1 =  [J_3,\bar J_2] + [J_2,\bar J_3] + [N,\bar J_1] - [J_1,\hat N] 
&&\nabla\bar J_1 = [N,\bar J_1] - [J_1,\hat N] \cr \cr
\bar \nabla J_2 =  [J_3,\bar J_3]  + [N,\bar J_2] - [J_2,\hat N] 
&&\nabla\bar J_2 = - [J_1,\bar J_1] + [N,\bar J_2] - [J_2,\hat N] \cr\cr
\bar \nabla J_3 = [N,\bar J_3] - [J_3,\hat N] 
&&\nabla\bar J_3 =- [J_1,\bar J_2] - [J_2,\bar J_1]  + [N,\bar J_3] - [J_3,\hat N] \cr\cr
\bar \nabla N = -[N,\hat N]  
&&\nabla \hat N =   [N,\hat N] 
\end{array}
\ee


\subsection{The current algebra}\label{subKalgebra}

In this section we discuss the current-current OPEs that are the elementary input needed for the computations of section \ref{fusion}.
The set of currents we consider are the currents $J_0$, $J_1$, $J_2$, $J_3$ as well as the ghost Lorentz current $N$, together with their anti-holomorphic partners  $\bar J_0$, $\bar J_1$, $\bar J_2$, $\bar J_3$ and $\hat N$. In order to simplify the expressions in the following computations, we introduce the generic notation $K_m$, $\bar K_m$ for the currents. The index $m$ takes the values in the set $\{ 0,1,2,3,g \}$.
For $m=0,1,2,3$ we define $K_m\equiv J_m$, $\bar K_m \equiv \bar J_m$.
For the particular value $m=g$ we define $K_g\equiv N$, $\bar K_g \equiv \hat N$, and $g$ stands for ``ghost".
The index $m$ codes the $\mathbb{Z}_4$-grade of the current. The ghost currents have grade zero.

The OPEs of the gauge covariant currents have been discussed in various papers. The OPEs at first-order in the $R^{-2}$ expansion have been analyzed in \cite{Bianchi:2006im}\cite{Puletti:2006vb}\cite{Puletti:2008ym}\cite{Mikhailov:2007mr}. The $R^{-4}$ corrections to the second-order poles have been computed in \cite{Bedoya:2010av}. 
For the purpose of the present article, the knowledge of the current algebra at order $R^{-2}$ is enough. 

In appendix \ref{covCurrents} we present a new and rather efficient way of computing the current algebra. 
The idea is to demand compatibility with the Maurer-Cartan equation and with the equations of motion (more precisely with the reparametrization invariance of the path integral).
Notice that the OPEs involving $J_0$ and $\bar J_0$ generically suffer from some ambiguities because of the gauge freedom. 
At the end of appendix \ref{covCurrents} we compare the version of the current algebra we compute with the ones that appeared previously in the literature.

It is convenient to expand the current on a basis of the Lie superalgebra that is compatible with the $\mathbb{Z}_4$ grading. We write:
\be K_m = K_m^{A_m} t_{A_m} \ee
The indices $A_m$ are adjoint indices\footnote{In most of the literature the notation are $A_0\to [ab]$, $A_1 \to \alpha$, $A_2 \to a$, $A_3 \to \hat \alpha$. Although slightly less explicit, the notations used here allow for a much more compact writing.} restricted to the subspace of the Lie superalgebra of grade $m$.
The generators $t_{A_m}$ form a basis of the subspace $\mathcal{H}_m$.

The current algebra takes the form:
\begin{align} \label{Kalgebra}
K_m^{A_m}(z) K_n^{B_n}(w)  = &
R^{-2} C_{mn} \frac{\kappa^{B_n A_m}}{(z-w)^2}
+ R^{-2} \sum_p C_{mn}^p \frac{{f_{C_p}}^{B_n A_m} K_p^{C_p}}{z-w} \cr
& + R^{-2} \sum_p C_{mn}^{\bar p} {f_{C_p}}^{B_n A_m} \bar K_p^{C_p} \frac{\bar z - \bar w}{(z-w)^2} +... \cr
K_m^{A_m}(z) \bar K_n^{B_n}(w)  = &
R^{-2} C_{m \bar n} \kappa^{B_n A_m} 2\pi \delta^{(2)}(z-w)
+ R^{-2} \sum_p C_{m\bar n}^p \frac{{f_{C_p}}^{B_n A_m} K_p^{C_p}}{\bar z-\bar w} \cr
& + R^{-2} \sum_p C_{m\bar n}^{\bar p} \frac{{f_{C_p}}^{B_n A_m} \bar K_p^{C_p}}{z-w} +...\cr
\bar K_m^{A_m}(z) \bar K_n^{B_n}(w)  = &
R^{-2} C_{\bar m \bar n} \frac{\kappa^{B_n A_m} }{(\bar z - \bar w)^2}
+ R^{-2} \sum_p C_{\bar m\bar n}^p {f_{C_p}}^{B_n A_m} K_p^{C_p}\frac{z-w}{(\bar z-\bar w)^2} \cr
& + R^{-2} \sum_p C_{\bar m\bar n}^{\bar p} \frac{{f_{C_p}}^{B_n A_m} \bar K_p^{C_p}}{\bar z-\bar w} +...
\end{align}
The tensors $\kappa^{AB}$ and ${f_{C}}^{BA}$ are respectively the metric and the structure constants (see appendix \ref{conventions} for conventions).
The non-trivial data in the current algebra  \eqref{Kalgebra} is coded in the coefficients $C_{**}$, $C_{**}^{*}$. 
These coefficients should be read as follows: the coefficient $C_{13}$ give the coefficient of the identity operator in the OPE between the currents $J_1$ and $J_3$, the coefficient $C_{2\bar 2}^g$ give the coefficient of the ghost current $N$ in the OPE between $J_2$ and $\bar J_2$, and so on.
We introduced a sum over all currents in the first-order poles in order to simplify the writing, but many of the $C$'s are clearly zero since they do not respect the $\mathbb{Z}_4$ grading.
The coefficients $C_{**}$, $C_{**}^{*}$ are symmetric in their two lower indices.
Also $\sum_p C_{mn}^{\bar p}$ should be understood as $C_{mn}^{\bar 1} + C_{mn}^{\bar 2} +... $.
The non-zero coefficients are given below.
The non-vanishing second-order poles are:
\be C_{13} = -1 ,\quad C_{1 \bar 3} = 1 ,\quad C_{\bar 1 3} = 1 ,\quad C_{\bar 1 \bar 3} = -1 ,\quad C_{22} = -1 ,\quad C_{2\bar 2} = 1 ,\quad C_{\bar 2 \bar 2} = -1 \ee
The first-order poles involving only the currents of non-zero grade are:
\be \begin{array}{c}
C_{11}^2 = 2 ,\quad C_{11}^{\bar 2} = 1 ,\quad C_{1 \bar 1}^2 = 1 ,\quad C_{\bar 1 \bar 1}^{\bar 2} = 1 \cr\cr
C_{33}^2 = 1 ,\quad C_{3\bar 3}^{\bar 2} = 1 ,\quad C_{\bar 3 \bar 3}^2 = 1 ,\quad C_{\bar 3 \bar 3}^{\bar 2} = 2 \cr\cr
C_{12}^3 = 2 ,\quad C_{12}^{\bar 3} = 1 ,\quad C_{\bar 1 2}^3 = 1 ,\quad C_{1\bar 2}^3 = 1 ,\quad C_{\bar 1 \bar 2}^{\bar 3} = 1 \cr\cr
C_{32}^1 = 1 ,\quad C_{\bar 3 2}^{\bar 1} = 1 ,\quad C_{3 \bar 2}^{\bar 1} = 1 ,\quad C_{\bar 3 \bar 2}^1 = 1 ,\quad C_{\bar 3 \bar 2}^{\bar 1} =2 
\end{array} \ee  
The first-order poles involving the ghosts currents are:
\be \begin{array}{c}
C_{22}^g = 1 ,\quad C_{22}^{\bar g} = -1 ,\quad C_{2 \bar 2}^g = 1 ,\quad C_{2 \bar 2}^{\bar g} = 1 ,\quad C_{\bar 2 \bar 2}^g = -1 ,\quad C_{\bar 2 \bar 2}^{\bar g} = 1 \cr\cr
C_{13}^g = 1 ,\quad C_{13}^{\bar g} = -1 ,\quad C_{1 \bar 3}^g = 1 ,\quad C_{1 \bar 3}^{\bar g}= 1 ,\quad C_{\bar 1 3}^g = 1 ,\quad C_{\bar 1 3}^{\bar g} = 1 ,\quad C_{\bar 1 \bar 3}^g = -1 ,\quad C_{\bar 1 \bar 3}^{\bar g} = 1 \cr\cr
C_{gg}^g = -1 ,\quad C_{\bar g \bar g}^{\bar g} = -1
\end{array} \ee 
Eventually the first-order poles involving the currents $J_0$, $\bar J_0$ are:
\be \begin{array}{c}
C_{10}^1 = 1 ,\quad C_{0 \bar 1}^1 = 1 ,\quad C_{\bar 0 1}^{\bar 1} = 1 ,\quad C_{\bar 0 \bar 1}^{\bar 1} = 1 \cr\cr
C_{02}^2 = 1 ,\quad C_{0 \bar 2}^2 = 1 ,\quad C_{\bar 0 2}^{\bar 2} = 1 ,\quad C_{\bar 0 \bar 2}^{\bar 2} = 1 \cr\cr
C_{30}^3 = 1 ,\quad C_{0 \bar 3}^3 = 1 ,\quad C_{\bar 0 3}^{\bar 3} = 1 ,\quad C_{\bar 0 \bar 3}^{\bar 3} = 1 \cr\cr
C_{13}^0 = 1 ,\quad C_{13}^{\bar 0} = 1 ,\quad C_{\bar 1 \bar 3}^0 = 1 ,\quad C_{\bar 1 \bar 3}^{\bar 0} = 1 \cr\cr
C_{22}^0 = 1 ,\quad C_{22}^{\bar 0} = 1 ,\quad C_{\bar 2 \bar 2}^0 = 1 ,\quad C_{\bar 2 \bar 2}^{\bar 0} = 1
\end{array}
\ee
The method we are using to compute the current algebra does not fix completely the self-OPEs of $J_0$ and $\bar J_0$. We only obtain the following constraints:
\begin{align}\label{constraintsC00*}
C_{00}^0 = C_{0\bar 0}^0 = -C_{\bar 0 \bar 0}^0 \quad ; \quad 
- C_{00}^{\bar 0} = C_{0\bar 0}^{\bar 0} = C_{\bar 0 \bar 0}^{\bar 0} \cr
C_{00}^g = C_{0\bar 0}^g = -C_{\bar 0 \bar 0}^g \quad ; \quad 
- C_{00}^{\bar g} = C_{0\bar 0}^{\bar g} = C_{\bar 0 \bar 0}^{\bar g} 
\end{align}
It turns out that these constraints are enough to perform explicitly the computations presented in this paper\footnote{More precisely, the constraints \eqref{constraintsC00*} implies that the coefficients cancel against each other in the computation of the commutator of equal-times connections \eqref{r,sSyst}. The reason is essentially that the currents $J_0$, $\bar J_0$ appear in the flat connection \eqref{defA} with no dependence on the spectral parameter. This in turns implies that these coefficients cancel against each other in the computation of the fusion line operators presented in section \ref{fusion}.
Notice however that the cancellation of some divergences in the line operators depends on the value of the coefficients \eqref{constraintsC00*}, see \eqref{possibleValueC00*}.}.


\subsection{The flat connection and the $(r,s)$ system}
Similarly to the Green-Schwarz string \cite{Metsaev:1998it}\cite{Bena:2003wd}, the pure spinor string on $AdS_5 \times S^5$ admits a one-parameter family of flat connections \cite{Vallilo:2003nx}.
This implies that the classical theory admits an infinite number of conserved charges.
The flat connection $A(y)$ is defined as:
\begin{align}\label{defA} A(y) = & (J_0 + y J_1 + y^2 J_2 + y^3 J_3 + (y^4-1) N)dz \cr
& + (\bar J_0 + y^{-3} \bar J_1 + y^{-2} \bar J_2 + y^{-1} \bar J_3 + (y^{-4}-1)\hat N) d \bar z \end{align}
The flat connection is invariant under parity combined with the exchange of $y$ and $y^{-1}$.
The equations of motion together with the Maurer-Cartan equation \eqref{MC+EOM} imply that the previous connection is flat for all values of the spectral parameter $y$:
\be d A(y) + A(y) \wedge A(y) = 0 \ee
In the following we study line operators that are the path-ordered exponential of the integral of the flat connection along a given contour.
We will only consider integration contours that lie at constant time. Consequently only the spacelike component of the flat connection will appear.
For simplicity, we use the same notation $A(y)$ for the connection and for its spacelike component.

The advantage of the version of the current algebra we are working with is that the commutator of two equal-time space-component of the flat connection can be written as a $(r,s)$ system:
\begin{align}\label{r,sSyst}
[ A_R(y;\sigma), A_{R'}(y';\sigma')] = &2s \p_\sigma\delta^{(2)}(\sigma-\sigma') + [ A_R(y;\sigma)+ A_{R'}(y';\sigma'),r] \delta^{(2)}(\sigma-\sigma') \cr
& + [ A_R(y;\sigma)- A_{R'}(y';\sigma'),s] \delta^{(2)}(\sigma-\sigma')
\end{align}
where $R$ and $R'$ denote the representations the two connections are transforming in.
The commutator transforms in the tensor product $R \otimes R'$.
Only the terms explicitly written down in the OPEs \eqref{Kalgebra} contribute to the commutator \eqref{r,sSyst}. The infinite number of subleading singularities contained in the ellipses of \eqref{Kalgebra} do not contribute to the commutator of equal-time currents (see e.g. \cite{Benichou:2010ts}).
As shown in appendix \ref{appRSmatrices}, the constant matrices $r$ and $s$ are given by:
\begin{align}\label{rMatrix}
 r = & i\pi R^{-2} \left( r_{13}t_{A_1}^R \otimes t_{B_3}^{R'} \kappa^{B_3 A_1} + r_{22}t_{A_2}^R  \otimes t_{B_2}^{R'} \kappa^{B_2 A_2} + r_{31}t_{A_3}^R \otimes  t_{B_1}^{R'} \kappa^{B_1 A_3} + r_{00}t_{A_0}^R  \otimes t_{B_0}^{R'} \kappa^{B_0 A_0} \right) \cr \cr
&r_{13} = \frac{(y^2-y^{-2})^2 + (y'^2-y'^{-2})^2}{y^4-y'^4}y y'^3 \quad ; \quad
r_{22} = \frac{(y^2-y^{-2})^2 + (y'^2-y'^{-2})^2}{y^4-y'^4}y^2 y'^2 \cr\cr
&r_{31} = \frac{(y^2-y^{-2})^2 + (y'^2-y'^{-2})^2}{y^4-y'^4}y^3 y' \quad ; \quad
r_{00} = 2 \frac{(y^4-1)(y'^4-1)}{y^4-y'^4} \end{align}
and:
\begin{align}\label{sMatrix}
 s = & i\pi R^{-2} \left( s_{13}t_{A_1}^R \otimes  t_{B_3}^{R'} \kappa^{B_3 A_1} + s_{22}t_{A_2}^R  \otimes t_{B_2}^{R'} \kappa^{B_2 A_2} + s_{31}t_{A_3}^R  \otimes t_{B_1}^{R'} \kappa^{B_1 A_3} + s_{00}t_{A_0}^R  \otimes t_{B_0}^{R'} \kappa^{B_0 A_0} \right) \cr\cr
& s_{13} = \frac{1}{y^3 y'} - y y'^3 \quad ; \quad
s_{22} = \frac{1}{y^2 y'^2} - y^2 y'^2 \quad ; \quad
s_{31} = \frac{1}{y y'^3} - y^3 y' \quad ; \quad
s_{00} = 0
\end{align}
Later it will be important that the $r$ matrix simplifies in the limit where the difference between the spectral parameters $y$ and $y'$ is small:
\be\label{rLim} y-y' \to 0\quad  \Rightarrow \quad r \sim \frac{i\pi R^{-2}}{2(y-y')} y(y^2+y^{-2})^2 t_A^R  \otimes t_B^{R'} \kappa^{BA} \ee
The $(r,s)$ matrices \eqref{rMatrix}, \eqref{sMatrix} first appeared in \cite{Mikhailov:2007eg}.
A detailed study of the $(r,s)$ system for string theory in $AdS_5\times S^5$ and its properties can be found in \cite{Magro:2008dv}\cite{Vicedo:2009sn}\cite{Vicedo:2010qd} (see appendix \ref{appCompKAlgebras} for more details).


\subsection{The line operators}

\paragraph{Definitions.}
We are interested in studying line operators that are the path-ordered exponential of the integral of the flat connection on a given contour. When the contour is an interval $[a,b]$, the line operator is called the transition matrix. We denote it as $T^{b,a}_R(y)$:
\be T^{b,a}_R(y) = P \exp \left( - \int_a^b A_R(y)\right) \ee
The transition matrix is labelled by the representation $R$ in which the flat connection transforms.
Flatness of the connection implies that the classical transition matrix does not depend on the integration path chosen. 
This property has been argued to extend to the quantum theory in \cite{Puletti:2008ym}.
For simplicity we consider only constant-time contours.

For string theory purposes we are lead to define the theory on a cylinder.
Then we can define the monodromy matrix which is the line operator associated with a closed contour winding once around the cylinder:
\be \Omega_R(y) = P \exp \left(-\oint A_R(y) \right) \ee
Flatness of the connection implies that the eigenvalues of the monodromy matrix are independent on time.
Consequently they code an infinite number of conserved charges.
Eventually the transfer matrix is the supertrace of the monodromy matrix:
\be\label{defTransfer} \mathcal{T}_R(y) = STr \ P \exp \left(-\oint A_R(y) \right) \ee

\paragraph{Regularization of UV divergences.}
In a quantum theory the line operators are generically ill-defined since the collisions of integrated connections lead to divergences. To properly define line operators one has to regularize these divergences, and then renormalize the line operators.
In order to study the UV divergences, we first have to expand the exponentials in the line operators. We write the transition matrix as:
\be T^{b,a}_R(y) = \sum_{M=0}^\infty (-1)^M T_{R,(M)}^{b,a}(y) \ee
where the $M$-th term is the path-ordered integral of $M$ connections:
\be T_{R,(M)}^{b,a}(y) = \frac{1}{M!}P\left(\int_a^b A_R(y) \right)^M = \int_{b>\sigma_1>...>\sigma_M>a}d\sigma_1...d\sigma_M A_R(y;\sigma_1)...A_R(y;\sigma_M) \ee
and similarly for the monodromy and transfer matrices.
Divergences occur when two integrated connections collide. It is clear from the current algebra \eqref{Kalgebra} that the collision of two connections leads to second- and first-order poles%
\footnote{The ellipses in the current algebra \eqref{Kalgebra} contain subleading singularities, including possible logarithmic singularities. Such terms do not lead to any UV divergences in the line operators. Indeed the integral of these subleading singularities gives a finite result.}.
In order to regularize these divergences, we introduce a UV cut-off $\epsilon$. We use a principal-value regularization scheme as suggested in \cite{Mikhailov:2007eg}. The OPE between two equal-time connections $A(\sigma)$ and $A(\sigma')$ is regularized by a small shift in time, in a symmetric way:
\be\label{defReg} A(\sigma) A(\sigma') \to \frac{1}{2} \left( A(\sigma+i\epsilon) A(\sigma') +  A(\sigma) A(\sigma'+i\epsilon) \right)\ee
For instance, a first-order pole is regularized as:
\be\label{reg1pole} \frac{1}{\sigma-\sigma'} \to P.V.\frac{1}{\sigma-\sigma'} = \frac{1}{2}\left(\frac{1}{\sigma+i\epsilon-\sigma'} + \frac{1}{\sigma-i\epsilon-\sigma'}\right) = \frac{\sigma-\sigma'}{(\sigma-\sigma')^2 + \epsilon^2} \ee
This regularization scheme turns out to be very convenient to discuss the fusion of line operators, as explained in section \ref{fusion}.

\paragraph{Divergences at order $R^{-2}$.}
The first-order divergences in line operators in the pure spinor string on $AdS_5 \times S^5$ were first studied in \cite{Mikhailov:2007mr}.
In this paper the authors used a different regularization scheme: the OPEs were regularized by imposing that the distance between two connections cannot be smaller than the UV cut-off.
The authors of \cite{Mikhailov:2007mr} also used a slightly different version of the current algebra.
In appendix \ref{computationDiv} we revisit the analysis of \cite{Mikhailov:2007mr} using the regularization scheme \eqref{defReg} and the current algebra \eqref{Kalgebra}.
The main difference we obtain with respect to \cite{Mikhailov:2007mr} is that the linear divergences do cancel thanks to our choice of regularization scheme.
Below we summarize the results derived in appendix  \ref{computationDiv} .

The transition matrices contain logarithmic divergences.
Schematically, these divergences read:
\be \sim  \log \epsilon \sum_{i=0}^3 \# \{ t^{A_i} t_{A_i}, T^{b,a}(y) \} \ee
The precise expression for these divergences is given in equation \eqref{divTransition}.
Consequently the transition matrices need to be renormalized. These divergences are cancelled by a simple wave-function renormalization.

The monodromy matrix also contains logarithmic divergences. These are given in \eqref{divMonodromy}. Schematically, these divergences read:
\be\label{divOmega} \sim  \log \epsilon \sum_{i=0}^3 \# \left( t^{A_i} t_{A_i} \Omega(y) + \Omega(y) t^{A_i} t_{A_i} - 2 t^{A_i} \Omega(y) t_{A_i}  \right) \ee
The new divergences with respect to the transition matrices come from collisions between connections sitting on both sides of the starting point of the integration contour.
These divergences are cancelled by a simple wave-function renormalization of the monodromy matrix.

The most important result for the purpose of this paper is that the transfer matrix is completely free of divergences at order $R^{-2}$.
This follows simply by taking the supertrace of equation \eqref{divOmega}.
This remarkable property strongly relies on the vanishing of the dual Coxeter number of the global symmetry group $PSU(2,2|4)$. For instance in generic WZW models, cancellation of divergences in the transfer matrices require both a wave-function renormalization and a renormalization of the spectral parameter \cite{Bachas:2004sy}. The divergences identified in \cite{Bachas:2004sy} also vanish if the dual Coxeter number of the group is zero.


\section{Fusion of line operators}\label{fusion}

In this section we study the fusion of two line operators. 
The fusion is the process of bringing the integration contours of two line operators on top of each other.
We are interested in the quantum effects that occur in this process.
The fusion of line operators for the pure-spinor string in $AdS_5 \times S^5$
was studied at first-order in perturbation theory in \cite{Mikhailov:2007eg}. In this section we will revisit and extend the first-order computations of \cite{Mikhailov:2007eg}. Then we will further extend the computation of fusion up to second order in perturbation theory.

The structure of the computations is similar to the ones presented in \cite{Benichou:2010ts}, where more details can be found.
In \cite{Benichou:2010ts} the computations were performed in the sigma model on the supergroup $PSl(n|n)$.
This theory is a good toy model for the pure spinor string on $AdS_5 \times S^5$. Indeed the complications coming from the coset structure and the pure spinor ghosts are absent.


\subsection{Setting up the computation}\label{recipeFusion}
Let us consider two transition matrices $T^{b,a}_R(y)$ and $T^{d,c}_{R'}(y')$ that transform respectively in the representations $R$ and $R'$. The fusion of these two matrices transforms in the tensor product $R\otimes R'$. In the following we will omit the symbol $\otimes$ to lighten the formulas. 
We represent the fusion of these two transition matrices with the symbol $\F$.
This process is defined as:
\be\label{defFusion} T^{b,a}_R(y)\F T^{d,c}_{R'}(y') = \lim_{\epsilon \to 0^+} T^{b+i\epsilon,a+i\epsilon}_R(y)T^{d,c}_{R'}(y') \ee
Assuming the integration contour of the transition matrices $T^{d,c}_{R'}(y')$ lies at constant time $\tau$, then the integration contour of $T^{b+i\epsilon,a+i\epsilon}_R(y)$ lies at constant time $\tau + \epsilon$.
If the intervals $[a,b]$ and $[c,d]$ do not overlap, the process of fusion is trivial.
In the following we assume that the overlap of these intervals is non-zero.
As the distance between the two contours goes to zero, the  OPEs between integrated connections sitting on the two contours produce quantum corrections to the classical process of fusion. These are the corrections we will evaluate.

Let us consider the OPE between two connections $A_R(y;\sigma+i\epsilon)$ and $A_{R'}(y';\sigma')$ integrated respectively on the first and on the second contour.
We write this OPE as:
\begin{align}\label{splitAA'}
A_R(y;\sigma+i\epsilon)A_{R'}(y';\sigma') = &\frac{1}{2}\left( A_R(y;\sigma+i\epsilon)A_{R'}(y';\sigma') +A_R(y;\sigma)A_{R'}(y';\sigma'+i\epsilon) \right) \cr &+  \frac{1}{2}\left( A_R(y;\sigma+i\epsilon)A_{R'}(y';\sigma') -A_R(y;\sigma)A_{R'}(y';\sigma'+i\epsilon) \right) \end{align}
Comparing with equation \eqref{defReg}, we notice that the first term in the previous equation should be understood as regularized OPE in the quantum line operator obtained after the fusion has been completed.
On the other hand, the second term in \eqref{splitAA'} should be understood as producing a quantum correction proper to the process of fusion. This are the corrections we want to compute.

In order to understand better the meaning of \eqref{splitAA'}, let us isolate a first-order pole in the OPE between the two connections.
Under the decomposition \eqref{splitAA'}, it is rewritten as:
\begin{align}\label{splitPole}  \frac{1}{\sigma+i \epsilon - \sigma'} 
& =  \frac{1}{2} \left( \frac{1}{\sigma+i \epsilon - \sigma'} +  \frac{1}{\sigma-i \epsilon - \sigma'} \right)
+ \frac{1}{2} \left( \frac{1}{\sigma+i \epsilon - \sigma'} -  \frac{1}{\sigma-i \epsilon - \sigma'} \right) \cr
& = P.V.\ \frac{1}{\sigma - \sigma'} - i \pi \delta_\epsilon(\sigma-\sigma')
\end{align}
The term we focus on is the second term on the right-hand side. As the notation suggests, it is actually a regularization of the delta-function.
Once we integrate upon the free coordinates, it produces a finite quantum corrections to the fusion of line operators. 
We can perform a similar manipulation for all first- and second-order poles appearing in the OPE between the two connections $A_R(y;\sigma+i\epsilon)$ and $A_{R'}(y';\sigma')$.
In order to isolate the contribution to the quantum corrections associated with fusion, we subtract the principal value from the singularities.
We obtain that all the terms that contribute to the quantum corrections from fusion come with (derivatives of) regularized delta functions.

The upshot is the following: in order to compute the quantum corrections in the process of fusion, we have to subtract the ``principal value'' piece from the OPE between connections.
What we are left with is essentially the commutator between connections that we computed in \eqref{r,sSyst}:
 \be\label{1-PVAA'}  (1-P.V.) A_R(y;\sigma+i\epsilon) A_{R'}(y';\sigma')  \overset{\epsilon \to 0^+}{=}  \frac{1}{2} [A_{R}(y;\sigma), A_{R'}(y';\sigma') ] \ee


\subsection{Fusion at first-order}
We begin with the corrections of order $R^{-2}$.
Since all terms in the current algebra are of order $R^{-2}$, it is enough to perform one OPE.
The computation of the first-order corrections in the fusion of two transition matrices was performed in \cite{Benichou:2010ts} using OPE techniques. This computation holds provided the commutator of connections can be written as a $(r,s)$ system, which is the case for the pure spinor string on $AdS_5\times S^5$ (see \eqref{r,sSyst}). The result obtained in \cite{Benichou:2010ts} matches the hamiltonian analysis of \cite{Maillet:1985ek}:
\begin{align}\label{fusionOrder1}
&  T^{b,a}_R(y)\F T^{d,c}_{R'}(y')
 =  T^{b,a}_R(y)T^{d,c}_{R'}(y') \cr
&+ \chi(b;c,d) T^{d,b}_{R'}(y') \frac{r+s}{2} T^{b,a}_R(y)T^{b,c}_{R'}(y') - \chi(a;c,d) T^{b,a}_R(y) T^{d,a}_{R'}(y') \frac{r+s}{2} T^{b,c}_{R'}(y') \cr
& + \chi(d;a,b) T^{b,d}_R(y) \frac{r-s}{2} T^{d,a}_R(y) T^{d,c}_{R'}(y') - \chi(c;a,b) T^{b,c}_R(y) T^{d,c}_{R'}(y') \frac{r-s}{2} T^{c,a}_R(y) \cr
& + \mathcal{O}(R^{-4})
\end{align}
The first term on the right-hand side is the zeroth-order result. The remaining terms are the first-order corrections. The function $\chi(a;b,c)$ is the characteristic function of the interval $[b,c]$ which takes the value $1$ if $b>a>c$ and $0$ if $a>b$ or $a<c$. For the special case where the integration intervals of the line operators have coinciding endpoints, that is for $a=b$ or $a=c$, then the characteristic function $\chi(a;b,c)$ has to be evaluated as $\frac{1}{2}$. This prescription essentially had to be guessed in the hamiltonian formalism \cite{Maillet:1985fn}.
In the OPE formalism it is a consequence of the definition \eqref{defFusion} \cite{Benichou:2010ts}.

Notice that the first-order corrections in \eqref{fusionOrder1} are anti-symmetric in the exchange of the two line operators. So they contribute only to the commutator of the line operators.
This follows from the fact that the quantum corrections associated with fusion come from the anti-symmetric part in the OPEs (see \eqref{1-PVAA'}).

From equation \eqref{fusionOrder1} we can deduce the fusion of transfer matrices \cite{Benichou:2010ts}.
We obtain that the fusion of transfer matrices at first order is trivial:
\begin{align}
\T_R(y) \F \T_{R'}(y') = \T_R(y) \T_{R'}(y') + \mathcal{O}(R^{-4})
\end{align}
Indeed the first-order corrections associated with fusion in  \eqref{fusionOrder1} take the form of constant matrices inserted at the endpoints of the overlap of the integration intervals. Since the transfer matrices have no endpoints, it is not surprising that these corrections vanish.
This implies in particular that the commutator of the transfer matrices is zero at first order.

\subsection{Fusion at second-order}\label{subFusion2nd}
Remember that our main goal is to show that the leading quantum correction in the fusion of two transfer matrices gives the shifts in the T-system at first order.
Since the fusion of transfer matrices is trivial at first-order, we need to study the fusion of line operators at second-order.

There are two different ways we can obtain $R^{-4}$ corrections in the process of fusion:
\begin{itemize}
\item The first way is to take one single OPE between two integrated connections, and include $R^{-4}$ corrections to the current algebra \eqref{Kalgebra}. 
As argued in section \ref{recipeFusion}, only the anti-symmetric part of the current-current OPEs contributes to the process of fusion.
Consequently at this order the $R^{-4}$ corrections to the current algebra lead to  $R^{-4}$ corrections to the commutator of line operators.
We postpone the computation of these corrections for future work since the current algebra is only partially known at order $R^{-4}$ \cite{Bedoya:2010av}.
\item The second way is to perform two OPEs between integrated connections and use the current algebra at order $R^{-2}$. 
Following the logic explained in section \ref{recipeFusion}, we again consider only the anti-symmetric part in each OPE. 
Since we perform an even number of OPEs this time the result will be symmetric under the exchange of the two line operators.
More precisely we obtain a contribution to the symmetric product of the line operators of order $R^{-4}$. This are the terms that we will compute in this paper.
\end{itemize}

More generally, the arguments of section \ref{recipeFusion} imply that any quantum correction in the process of fusion that involve an even (respectively odd) number of OPEs would contribute only to the symmetric product (respectively commutator) of the line operators. Consequently the $R^{-4}$ corrections to the current algebra would contribute to the symmetric product of line operators at order $R^{-6}$ and higher.

Below we describe the different steps in the computation of the symmetric fusion of line operators at order $R^{-4}$.
In an attempt to keep this section readable, the technical details of the computation have been gathered in appendix \ref{comFusion}.
More details can also be found in \cite{Benichou:2010ts}.

\begin{figure}
\centering
\includegraphics[scale=0.50]{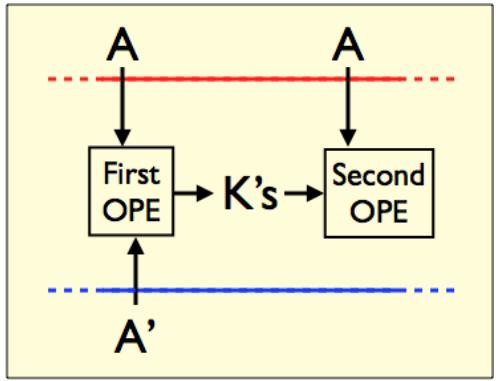}
\caption{A triple collision. The first OPE is taken between two connections sitting on different contours. The second OPE is taken between a third connection and the currents resulting from the first OPE.\label{tripleCollision}}
\end{figure}

Let us consider two line operators with contours separated in time by a small distance $\epsilon$.
We have to take two OPEs between the connections integrated on the contours.
The simplest way is to take two OPEs between two distinct pairs of connections. 
But we can also take one OPE between two connections, and then take the OPE of the resulting currents with a third connections. We will call this latter process a triple collision (see figure \ref{tripleCollision}).

\begin{figure}
\centering
\includegraphics[scale=0.55]{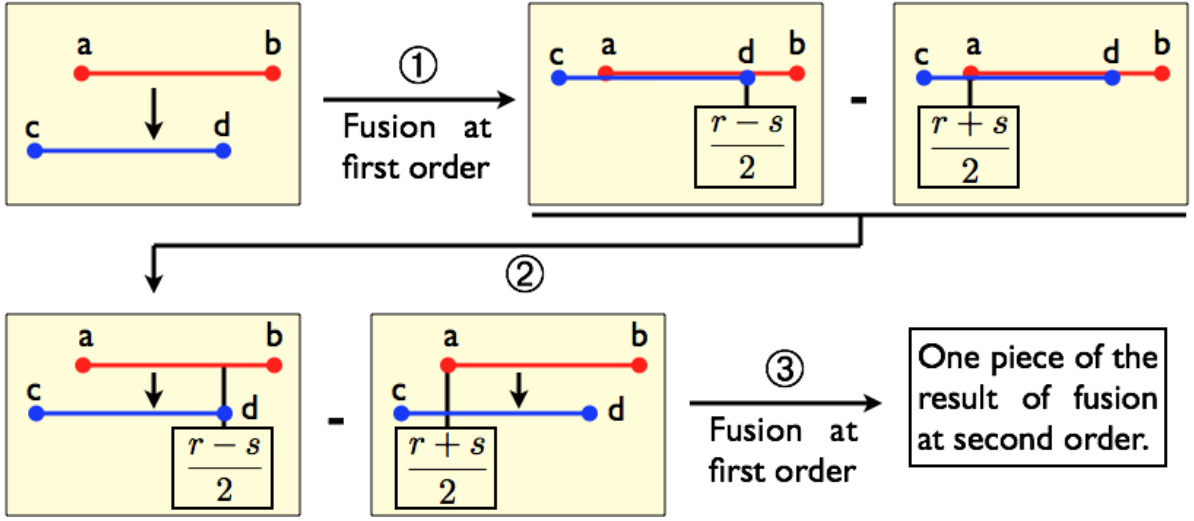}
\caption{We can compute one piece of the result for fusion at second order in the following way. In step \ding{172} we compute the first-order corrections from fusion. In step \ding{173} we separate the contours again. Finally in step \ding{174} we perform a second time the fusion at first order. This computation does not give the full result since the triple collisions are not properly accounted for. For transfer matrices the fusion at first order is trivial, so the quantum corrections obtained in this way actually vanish. \label{fusO2step1}}
\end{figure}

The computation is conveniently decomposed in two steps. 
\begin{itemize}
\item A first part of the total answer is obtained in the way depicted in Figure \ref{fusO2step1}.
We start from the result of the fusion at first order. We pull the contours away, and then re-fuse the line operators.
 The result of this procedure was computed in \cite{Benichou:2010ts} for an arbitrary $(r,s)$ system. 
We obtain new insertion of constant matrices at the endpoints of the overlap of the contours of the line operators.
 Roughly speaking, the first-order result \eqref{fusionOrder1} exponentiate\footnote{The precise expression is slightly more complicated than the one given in \cite{Benichou:2010ts}, since the formula (4.15) in \cite{Benichou:2010ts} does not generalizes to the coset. This implies that the exponentiation observed in \cite{Benichou:2010ts} is not exact in the case at hand.}.
What is important for our purposes is that this procedure gives once again a vanishing result for the fusion of transfer matrices. This simply follows from the fact that there is no first-order correction in the fusion of transfer matrices.
\item The procedure described above does not capture correctly the quantum corrections coming from triple collisions. 
Indeed in this procedure, the intermediate currents in the triple collisions are distributed in an arbitrary way on the two integration contours so that they recombine into connections.
This induces a source of errors.
So we have to compute separately additional corrections coming from the triple collisions.
This is done in appendix \ref{comFusion}.
\end{itemize}

The analysis of appendix \ref{comFusion} shows that there are two types of corrections that we need to add on top of the result obtained by the procedure described in figure \ref{fusO2step1}. 
The first type of corrective terms contain the integration of an operator $\tilde K$ on the overlap of the contours:
\begin{align}\label{FusionTT1}
& R^{-4} \sum_{M,M'=0}^\infty (-)^{M+M'}  \sum_{i=0}^M \sum_{i'=0}^{M'} \int_{[a,b]\cap[c,d]}d\sigma 
\left \lfloor \int_\sigma^b A \right\rceil^i \left \lfloor \int_\sigma^d A' \right\rceil^{i'}  \tilde K (\sigma)  \left \lfloor \int_a^\sigma A \right\rceil^{M-i} \left \lfloor \int_c^\sigma A' \right\rceil^{M'-i'} 
\end{align}
where we introduced the convenient notation $\lfloor \int_a^b A \rceil^M$ to describe the path-ordered integral of $M$ connections on the interval $[a,b]$.
The precise expression for the operator $\tilde K$ is explicitly given in \eqref{Ktilde}.
Schematically, the operator $\tilde K$ is a linear combination of the currents multiplied by three generators of the Lie superalgebra in the representations $R$ or $R'$, and contracted with structure constants.
The second type of corrective terms contain a constant matrix $\tilde{tt}$ inserted in between the integrated connections on the overlap of the integration contours:
\begin{align}\label{FusionTT2}
& R^{-4} \sum_{M,M'=0}^\infty (-)^{M+M'}  \sum_{i=0}^M \sum_{i'=0}^{M'} \int_a^b d\sigma \int_c^d d \sigma'
\left \lfloor \int_\sigma^b A \right\rceil^i \left \lfloor \int_\sigma^d A' \right\rceil^{i'} \cr
& \quad \times \delta^2_{\epsilon}(\sigma-\sigma') \tilde{tt}  \left \lfloor \int_a^\sigma A \right\rceil^{M-i} \left \lfloor \int_c^\sigma A' \right\rceil^{M'-i'} 
\end{align}
The precise expression for the matrix $\tilde{tt}$ is given in \eqref{tttilde}. Schematically, it is a linear combination of the tensor product of two generators taken in the representation $R$ and $R'$ and contracted with structure constants. Notice that the integration over the regularized delta function squared produce a linear divergence when the UV regulator $\epsilon$ is sent to zero. It would be interesting to perform a complete analysis of the second-order divergences in the line operators along the lines of section 3 in \cite{Benichou:2010ts}, to see whether the transfer matrices as well as the result of the fusion of transfer matrices are free of divergences up to second-order.

\begin{figure}
\centering
\includegraphics[scale=0.45]{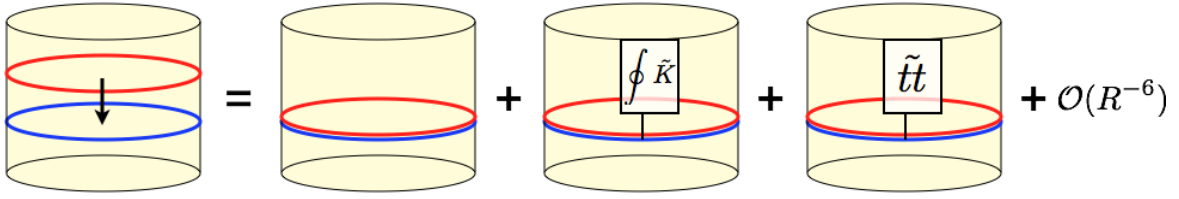}
\caption{Schematic representation of the symmetric fusion of transfer matrices at second order \eqref{summaryFusionTT}. The first term is the classical result. In the second term, an additional operator $\tilde{K}$ is integrated in between the connections. In the third term, a constant matrix $\tilde{tt}$ is inserted in between the integrated connections.\label{fusionTs}}
\end{figure}

\paragraph{Fusion of transfer matrices at second order: upshot.}
The detailed expression for the symmetric fusion of transition and monodromy matrices at second order is quite indigestible, so we refrain from giving an explicit formula for those. 
On the other hand the symmetric fusion of transfer matrices, that is crucial for the purposes of this paper, turns out to be rather simple:
\begin{align}\label{summaryFusionTT}
&\T_R(y)\ \SF \ \T_{R'}(y') = \frac{1}{2}\{ \T_R(y), \T_{R'}(y')\} \cr
& + R^{-4} STr \left(  \int_0^{2\pi} d\sigma \
T_R^{2\pi, \sigma}(y) T_{R'}^{2\pi, \sigma}(y') \ \tilde K (\sigma) \ T_R^{ \sigma,0}(y) T_{R'}^{\sigma,0}(y') \right)\cr
& + R^{-4} STr \left(  \int_0^{2\pi} d\sigma \int_0^{2\pi} d\sigma' \
T_R^{2\pi, \sigma}(y) T_{R'}^{2\pi, \sigma}(y') \delta^2_{\epsilon}(\sigma-\sigma')\ \tilde{tt}  \ T_R^{ \sigma,0}(y) T_{R'}^{\sigma,0}(y') \right)\cr
&+ \mathcal{O}(R^{-6})
\end{align}
where we denoted by $\SF$ the symmetrized fusion product.
This result is schematically represented in figure \ref{fusionTs}.
The detailed expressions for the operator $\tilde K$ and the constant matrix $\tilde{tt}$ can be read from equations \eqref{Ktilde} and \eqref{tttilde}.

\section{The $AdS_5$/$CFT_4$ T-system}\label{proofTsys}

In this section we use the previous computations to obtain a first-principle perturbative derivation of the T-system.
As explained in the introduction, the idea is to promote the T-system \eqref{Tsystem} to an operator identity, where the product between transfer matrices is understood as the fusion product:
\be\label{QTsystem} \mathcal{T}_{a,s}(u + 1) \F \mathcal{T}_{a,s}(u - 1) = 
  \mathcal{T}_{a+1,s}(u+1)\F \mathcal{T}_{a-1,s}(u-1) + \mathcal{T}_{a,s-1}(u+1) \F \mathcal{T}_{a,s+1}(u-1) \ee
We expect the transfer matrices to commute in the quantum theory. This has been proven in section \ref{fusion} at order $R^{-2}$. Consequently we can equivalently use the symmetric fusion product to define the T-system \eqref{QTsystem}.

The integer label $a,s$ label unitary irreducible representation of $PSU(2,2|4)$. These labels take value in a T-shaped lattice (see e.g. \cite{Gromov:2010kf}\cite{Volin:2010xz}). The corresponding representations are associated with rectangular Young tableaux which size is given by the value of the labels $a,s$.
It is known that these representations satisfy the following supercharacter identity (see e.g. \cite{Gromov:2010vb}\cite{Kazakov:2007na}):
\be\label{charId1} \chi(a,s)^2 = \chi(a+1,s)\chi(a-1,s)+\chi(a,s+1)\chi(a,s-1) \ee
In the limit where we neglect both the shifts of the spectral parameter as well as the quantum effects associated with fusion, the T-system \eqref{QTsystem} reduces to the character identity \eqref{charId1}.

Next we want to show that the T-system \eqref{QTsystem} holds at first order. We consider:
\begin{align}\label{Tsys1}
0  \stackrel{?}{=} & \mathcal{T}_{a,s}(y+\delta) \F \mathcal{T}_{a,s}(y-\delta) - 
  \mathcal{T}_{a+1,s}(y+\delta)\F \mathcal{T}_{a-1,s}(y-\delta) - \mathcal{T}_{a,s-1}(y+\delta) \F \mathcal{T}_{a,s+1}(y-\delta) \cr &  \equiv \sum_{R,R'} \mathcal{T}_{R}(y+\delta) \F \mathcal{T}_{R'}(y-\delta) \end{align}
where we use the shorthand $\sum_{R,R'}$ to denote the sum over representations that appears in the T-system.
We look for a value of the shift of the spectral parameters $\delta$ such that the previous quantity does indeed vanish, and the T-system holds. 
Then we can deduce the relationship between the spectral parameter $y$ used to define the flat connection \eqref{defA}, and the spectral parameter $u$ that appears in the T-system \eqref{QTsystem}.

We assume that $\delta$ is of order $R^{-2}$. We will now show that the terms of order $R^{-2}$ in \eqref{Tsys1} do vanish. More precisely, the terms of order $R^{-2}$ coming from the derivative expansion of the transfer matrices cancel against the leading quantum correction coming from the process of fusion.

\paragraph{Fusion of transfer matrices when the difference of spectral parameter is small.}
In section \ref{fusion} we obtained that the leading quantum correction in the fusion of transfer matrices is of order $R^{-4}$. However when the difference of the spectral parameter is of order $R^{-2}$, a piece of the leading quantum correction actually becomes of order $R^{-2}$. So it has the right order of magnitude to cancel the first term in the derivative expansion of the transfer matrices in \eqref{Tsys1}. More details can be found at the end of appendix \ref{comFusion}.

The interesting term comes from the second line in \eqref{summaryFusionTT}.
We assume $y-y' = \mathcal{O}(R^{-2})$.
From equations \eqref{KtildeLim} we obtain that \eqref{summaryFusionTT} simplifies to:
\begin{align}\label{fusTTSmallDif}
&\T_R(y) \F \T_{R'}(y') = \T_R(y) \T_{R'}(y') \cr
& + R^{-4} \frac{\pi^2 }{32}   \frac{y^2(y^2-y^{-2})^4}{y-y'} 
STr \left(  \int_0^{2\pi} d\sigma 
T_R^{2\pi, \sigma}(y) T_{R'}^{2\pi, \sigma}(y') 
\right. \cr &
\quad \times \left( -\sum_{m,n,p,q,r=0}^3\p_y A^{E_r}(y)
 \left({f_{C_p}}^{B_n A_m} {f_{E_r}}^{C_p D_q} \{ t_{D_q}^R , t_{A_m}^R] t_{B_n}^{R'} + {f_{C_p}}^{B_n A_m} {f_{E_r}}^{ D_q C_p}  t_{A_m}^R \{ t_{B_n}^{R'}, t_{D_q}^{R'}]\right)\right)\cr
&\quad \times \left. T_R^{ \sigma,0}(y) T_{R'}^{\sigma,0}(y') \right) + \mathcal{O}(R^{-4})
\end{align}
We observe that the linear combination of the currents that appears is now proportional to the derivative of the flat connection.

\paragraph{Useful character identities.}
To further simplify the quantum corrections that appear in \eqref{Tsys1}, we need to use some character identities that holds for the particular combination of representations that appear in the T-system.
In \cite{Kazakov:2007na} the validity of the T-system was proven for transfer matrices associated to $Gl(k|m)$ spin chains. 
Then a large family of character identities was deduced by expanding the T-system as an infinite series.
Here we go the other way: we try to reconstruct the T-system from a perturbative expansion.
Thus it makes sense that we need to use the character identities of \cite{Kazakov:2007na}.
In the appendix E of \cite{Benichou:2010ts}, it was shown that some of these character identities imply in particular that for any group element $g$ and for any function $K^E$:
\begin{align}\label{charIdKV1}
\sum_{R,R'} K^E {f_C}^{BA} {f_E}^{CD} & STr(\{ t_D^R, t_A^R] g^R \otimes t_B^{R'} g^{R'}) \cr
& = 2 \sum_{R,R'} STr( g^R \otimes K^E t_E^{R'} g^{R'}) - STr( K^E t_E^{R}  g^R \otimes g^{R'})  \end{align}
and similarly:
\begin{align}\label{charIdKV2}
\sum_{R,R'} K^E {f_C}^{BA} {f_E}^{DC} & STr(t_A^R g^R \otimes \{ t_B^{R'},t_D] g^{R'}) \cr
& = 2 \sum_{R,R'} STr( g^R \otimes K^E t_E^{R'} g^{R'}) - STr( K^E t_E^{R}  g^R \otimes g^{R'})  \end{align}
Essentially, these character identities allow to replace the complicated combination of structure constants and generators appearing in \eqref{fusTTSmallDif} by single generators, assuming we consider the sum of representation that appear in the T-system.
So we obtain:
\begin{align}\label{Tsys2}
\sum_{R,R'} & \T_R(y+\delta) \F  \T_{R'}(y-\delta) =  \sum_{R,R'}  \T_R(y+\delta) \T_{R'}(y-\delta) 
+  \sum_{R,R'} R^{-4}\frac{\pi^2 }{16}   \frac{y^2(y^2-y^{-2})^4}{\delta} \cr
& \times STr \left(  \int_0^{2\pi} d\sigma 
T_R^{2\pi, \sigma}(y) T_{R'}^{2\pi, \sigma}(y) \left( - \p_y A^{E}(y;\sigma) (-t_E^R + t_E^{R'}) \right)
T_R^{ \sigma,0}(y) T_{R'}^{\sigma,0}(y) \right)
+ \mathcal{O}(R^{-4})\cr
 =& \sum_{R,R'}  \T_R(y+\delta) \T_{R'}(y-\delta)
- R^{-4} \frac{\pi^2 }{16}   \frac{y^2(y^2-y^{-2})^4}{\delta} \left( \p_y \T_R(y) \T_{R'}(y) - \T_R(y) \p_y \T_{R'}(y) \right) \cr
&\quad + \mathcal{O}(R^{-4})
\end{align}

\paragraph{The T-system at first order.}
Performing a Taylor expansion for the $\T$'s in \eqref{Tsys2}, we deduce:
\begin{align}\label{Tsys3}
\sum_{R,R'} & \T_R(y+\delta) \F  \T_{R'}(y-\delta) =
 \sum_{R,R'}  \T_R(y) \T_{R'}(y) \cr
& +\left(\delta - R^{-4} \frac{\pi^2 }{16}   \frac{y^2(y^2-y^{-2})^4}{\delta} \right) \left( \p_y \T_R(y) \T_{R'}(y) - \T_R(y) \p_y \T_{R'}(y) \right) + \mathcal{O}(R^{-4})
\end{align}
The first term in the previous expression vanishes because of the classical identity \eqref{charId1}.
In order for the first-order corrections to vanish as well, we have to take:
\be\label{defDelta} \delta = R^{-2} \frac{\pi}{4}y(y^2-y^{-2})^2 \ee
Thus have have shown that:
\be\label{Tsys4} \sum_{R,R'}  \T_R(y+\delta) \F  \T_{R'}(y-\delta) = 0 + \mathcal{O}(R^{-4}) \ee

\paragraph{Redefinition of the spectral parameter.}
In order to write the T-system in the canonical form \eqref{QTsystem} we define:
\be\label{defU} u = \frac{R^2}{\pi} \frac{1}{1-y^4} + cst \ee
such that $u(y\pm \delta)=u(y) \pm 1$. 
Then equation \eqref{Tsys4} is rewritten as:
\be \mathcal{T}_{a,s}(u + 1) \F \mathcal{T}_{a,s}(u - 1) = 
  \mathcal{T}_{a+1,s}(u+1)\F \mathcal{T}_{a-1,s}(u-1) + \mathcal{T}_{a,s-1}(u+1) \F \mathcal{T}_{a,s+1}(u-1) + ... \ee
This is the canonical form of the T-system \eqref{Tsystem}.

\paragraph{Comparison with the T-system obtained via the Thermodynamic Bethe Ansatz.}
We can now perform a consistency check with the Thermodynamic Bethe Ansatz derivation of the $AdS_5\times S^5$ T-system  \cite{Gromov:2009bc}\cite{Bombardelli:2009ns}\cite{Arutyunov:2009ur}.
In this context the T-system is obtained in a slightly different form: only the T-functions on the left-hand side have a shifted spectral parameter:
\be\label{TsystemTBA} \mathcal{T}^{TBA}_{a,s}(u + 1)  \mathcal{T}^{TBA}_{a,s}(u - 1) = 
  \mathcal{T}^{TBA}_{a+1,s}(u) \mathcal{T}^{TBA}_{a-1,s}(u) + \mathcal{T}^{TBA}_{a,s-1}(u)  \mathcal{T}^{TBA}_{a,s+1}(u)  \ee
This mismatch is easily cured by a redefinition of the T-functions. Let us define the functions $\T^{TBA}(u)$ as: 
\be \T^{TBA}_{a,s}(u) = \T_{a,s}(u+a-s)\ee
Then the functions $\T^{TBA}$'s satisfy \eqref{TsystemTBA} if and only if the functions $\T$'s satisfy the T-system \eqref{Tsystem}.

However the previous redefinition does not change the magnitude of the shift on the left-hand side of the T-system. Thus the matching of the shifts gives a quantitative check of the consistency between the approach taken in this paper, and the TBA approach\footnote{The author would like to thank N. Gromov for stressing this point.}.
Next we perform this matching using the conventions of  \cite{Gromov:2009tv}\cite{Gromov:2009bc}.
The flat connection is written in terms of a spectral parameter $x$ as $A(x) = J_0 dz + (x-1)/(x+1)J_2 dz +...$. 
Comparing with \eqref{defA} we deduce that the spectral parameter $y$ that we use is related to $x$ as: $y^2=(x-1)/(x+1)$.
The variable $u^{TBA}$ that enters the TBA T-system is linked to the spectral parameter $x$ via the Zhukowsky map: $u^{TBA}/g=x+1/x$, where $g$ is related to the 't Hooft coupling $\lambda$ as $g=\sqrt{\lambda}/4\pi$.
The parameter $R$ can be linked to the 't Hooft coupling $\lambda$ by identification of the prefactor of the worldsheet action. This gives $\sqrt{\lambda}/2\pi = R^2/4\pi$.
Consequently the parameter $u$ \eqref{defU} that we obtained is related to the parameter $u^{TBA}$ as:
$ u = 2 u^{TBA}$,
assuming the free constant in \eqref{defU} takes the value $R^2/2\pi$.
The analysis of \cite{Gromov:2009bc} gives a T-system where the parameter $u^{TBA}$ is shifted by $\pm i/2$. Given that $u$ is shifted in our case by $\pm 1$, there is an apparent mismatch by a factor of $i$.
This comes from the fact that we have been working on an euclidean worldsheet. If we Wick-rotate the worldsheet to a minkowskian signature, then our analysis produce the T-system with imaginary shifts%
\footnote{Let us be a bit more precise on this point. On a minkowskian worldsheet, the imaginary shift $i \epsilon$ in e.g. \eqref{splitPole} would be replaced by a real shift $\epsilon$. This implies that in the computation of fusion all OPEs would come with an additional factor of $i$. Consequently the second-order corrections from fusion would come with an additional minus sign. This would in turn induce a factor of $i$ in the shift of the T-system.
} in perfect agreement with the TBA analysis.

\paragraph{Upshot.}
As claimed previously, we have derived the T-system up to first order in the large radius expansion.
More precisely, we have sown that the shifts of the spectral parameter in the T-system come from quantum effects in the fusion of transfer matrices.
Moreover we have checked that the shifts are the same than the ones obtained in the Thermodynamic Bethe Ansatz derivation of the T-system.

Notice that the vanishing of the divergences in the transfer matrices is important. Indeed if the transfer matrices would need to be renormalized, then the renormalization factor would most likely depend on the representation in which the transfer matrix is taken (see e.g. \cite{Bachas:2004sy}). It implies that the different terms in the T-system would be renormalized differently, which would most likely destroy the balance needed for the previous computation to work.


\section{Generalization to other integrable theories}\label{extensions}

In this paper we have proven that the T-system is realized in the pure spinor string on $AdS_5 \times S^5$ up to first order in the large radius expansion. 
In \cite{Benichou:2010ts} a similar proof was given for the non-linear sigma model on the supergroup $PSl(n|n)$. 
It is natural to look for other theories where this derivation can be easily generalized.
A close look at the computation leads to the following conclusion:
there are only a few necessary and sufficient conditions that a given model has to fulfill in order for the derivation to apply.
Obviously the theory has to exhibit a one-parameter family of flat connections. We assume that the connection takes value in a Lie algebra. 
The other conditions are the following:
\begin{itemize}
\item The equal-time commutator of the spacelike component of the connection can be written as a $(r,s)$ system.  
This guarantees that formula \eqref{summaryFusionTT} can be directly reproduced.
Moreover the $r$ matrix must satisfy a property similar to equation \eqref{rLim}: in the limit where the difference between the spectral parameter is small, the $r$ matrix needs to be proportional to the Casimir $\kappa^{BA} t_A \otimes t_B$. This condition is necessary for the simplification leading to equation \eqref{fusTTSmallDif} to occur.
\item The Lie group needs to possess an equivalent of the character identities \eqref{charIdKV1}, \eqref{charIdKV2}. 
Given the role played by the results of \cite{Kazakov:2007na}, it is tempting to speculate that a sufficient condition is that there exists a spin chain with the same symmetry group that realizes the T-system.
\item Eventually the transfer matrix has to be free of divergences at first order in perturbation theory\footnote{Actually a weaker condition is that all combinations of transfer matrices that enters the T-system are renormalized with the same coefficient.}. A crucial condition here is that the dual Coxeter number of the symmetry group vanishes. 
This condition prevents the renormalization of the transfer matrices from destroying the balance needed for the computation to work.
\end{itemize}
Next we discuss several candidate theories that may fulfill these requirements.

\paragraph{Candidate theories relevant for the AdS/CFT correspondence.}
Let us begin with the theories that describe superstrings in Anti-de Sitter backgrounds.
These theories are built on sigma models on (coset of) supergroups, see e.g. \cite{Zarembo:2010sg} for a classification of the relevant $\mathbb{Z}_4$ cosets. Notice that all the supergroups involved have a vanishing dual Coxeter number. This should not come as a surprise, since the vanishing of the dual Coxeter number is tightly related to the vanishing of the spacetime supercurvature, and thus to the fact that the equations of motion of supergravity are satisfied.

Spacetime covariance was helpful in the previous analysis. Consequently we will mostly discuss theories of the pure-spinor type that allow for a covariant quantization. Obviously it would be interesting to reproduce the previous computations in Green-Schwarz-like theories that realize kappa-symmetry. The structure of the computations would be identical, but the computations themselves would be more tedious because the gauge-fixing of kappa symmetry usually comes with a breaking of the target space isometries.

The first obvious candidate is string theory on $AdS_4 \times CP^3$. Indeed in \cite{Gromov:2009tv} a Y-system was conjectured to hold in this theory. The TBA derivation of the Y-system was performed in \cite{Bombardelli:2009xz}\cite{Gromov:2009at}.
In order to actually reproduce the computations described in the present paper, the pure spinor formulation of superstring theory in $AdS_4 \times CP^3$ developed in \cite{Fre:2008qc} is a natural starting point (see also \cite{Stefanski:2008ik}).

The second candidate is string theory on $AdS_3 \times S^3$.
The analysis of \cite{Benichou:2010ts} applies to the sigma-model on $PSU(1,1|2)$. In \cite{Berkovits:1999im} the hybrid formalism was developed to describe superstrings in $AdS_3 \times S^3$ in a superspace with eight supercharges.
In this formalism the worldsheet theory is the sigma model on $PSU(1,1|2)$, coupled to ghosts. This theory admits a consistent expansion in the ghosts. Thus the analysis of \cite{Benichou:2010ts}  implies that the hybrid string in $AdS_3 \times S^3$ realizes the T-system up to first order in the large radius expansion, and up to zeroth order in the ghosts expansion. 
This is valid for $AdS_3 \times S^3$ supported by RR fluxes, NS fluxes or by any mixing of these fluxes.
It would be instructive to dress up the computation of \cite{Benichou:2010ts} with the hybrid ghosts.
There has been some interest in the question of integrability for string theory in $AdS_3 \times S^3$, see e.g. \cite{Pakman:2009mi}. However the progress have been rather slower than in the case of $AdS_5 \times S^5$, mostly because the dual Conformal Field Theory is not as well understood. Presumably the approach presented in \cite{Benichou:2010ts} and in the present article can lead to a faster road to the solution of this problem.

Other formulations of string theory on $AdS_3 \times S^3$ involve a $\mathbb{Z}_4$ coset of the supergroup $PSU(1,1|2)\times PSU(1,1|2)$. It is the case of the hybrid string with sixteen manifest supercharges \cite{Berkovits:1999du}. It is reasonable to expect that the T-system is also realized in this formalism.

Another natural candidate is the hybrid description of superstrings in $AdS_2 \times S^2$ discussed in \cite{Berkovits:1999zq}. It is also based on a $\mathbb{Z}_4$ coset of the supergroup $PSU(1,1|2)$, and it is integrable \cite{Young:2005jv}\cite{Adam:2007ws}.

In the classification of \cite{Zarembo:2010sg} we find other candidate string backgrounds that are $\mathbb{Z}_4$ coset of supergroups with vanishing dual Coxeter number: $AdS_3\times S^3 \times S^3$, $AdS_2\times S^2 \times S^2$,  $AdS_2 \times S^3$ and $AdS_2$. Quantum integrability is likely to show up at least in some of these examples. No formalism has been proposed to covariantly quantize string theory in these backgrounds yet.

\paragraph{Other candidates.}
Other theories that may not be directly relevant for string theory presumably also realize the T-system in the way described in this paper. 
These are the sigma models on (cosets of) supergroups with vanishing dual Coxeter number, some of which play a role in condensed matter (see e.g. \cite{PS}\cite{Efetov:1983xg}\cite{Zirnbauer:1999ua}).

A first example is the sigma model on the supergroup $OSp(2n+2|2n)$. This model shares many of the remarkable properties of the sigma model on $PSl(n|n)$ \cite{Bershadsky:1999hk}, see e.g. \cite{Babichenko:2006uc}.

Next $\mathbb{Z}_2$ cosets of supergroups with vanishing dual Coxeter number are classically integrable \cite{Bena:2003wd}\cite{Babichenko:2006uc}. 
Some of them also display nice quantum features \cite{Candu:2010yg}.
In these model there is a current that is both flat and conserved.
Consequently the current algebra is very similar to the one found in sigma models on supergroups \cite{Ashok:2009xx}.
This follows from the generic method introduced in \cite{Benichou:2010rk} and used in appendix \ref{covCurrents} to compute the current algebra.

Classical integrability extends to $\mathbb{Z}_4$ and more generally to $\mathbb{Z}_m$ cosets \cite{Young:2005jv}. It would be interesting to understand these models better. In particular it may shed some new light on the question of the role of the pure spinor ghosts for quantum integrability of the pure spinor string in $AdS_5 \times S^5$.


\section{Conclusion}\label{conclusion}

\paragraph{Summary of the results.}
We have studied the fusion of line operators in the pure spinor string on $AdS_5 \times S^5$ up to second order in perturbation theory. 
We deduced that the pure spinor string on $AdS_5 \times S^5$ realizes the T-system as an operator identity, with the fusion product, up to first order in the large 't Hooft coupling expansion.
The quantum effects in the fusion of the transfer matrices give the shifts in the T-system.

\paragraph{Comparison with the Thermodynamic Bethe Ansatz.}
The T-system was previously derived using the TBA machinery \cite{Gromov:2009bc}\cite{Bombardelli:2009ns}\cite{Arutyunov:2009ur}.
Here we will compare the advantages of both approaches.

A weakness of the TBA is that it relies on several assumptions that are notoriously difficult to check.
In particular one has to assume quantum integrability to start with. Moreover the spectrum of excitations that contribute in the thermodynamic limit essentially has to be guessed through the string hypothesis. 
The approach of the current article has the big advantage of starting from first principles.

The other drawback of the TBA is that the derivation of the T-system only applies to the ground state. The fact that the same set of equations also codes the spectrum of excited states upon analytic continuation is essentially an empirical observation.
In this paper we have derived the T-system as an operator identity.
Thus there is no doubt that all states of the theory satisfy the T-system.

On the other hand, the approach we are using here is intrinsically perturbative. The computations needed to derive the T-system at first order were already quite heavy. It would take a lot of efforts to go to the next order.
The TBA approach is free of this limitation since it produces the full T-system in one go.

There are also some by-products of the TBA approach that were not reproduced in the present work. In particular the TBA gives a explicit formula to extract the spectrum from the T-functions. It also gives some informations about the analytic properties of these functions. It would be interesting to investigate these questions with the elementary techniques used in the present paper. We hope  to come back to these questions in future work.

The computation presented in this paper gives a very strong argument in favor of the validity of the T-system.
It is not a definite proof since it is perturbative.
However the previous discussion shows that the approach presented here is complementary with the TBA analysis.
Indeed the weak points of the TBA are the the strong points in our approach, and vice-versa.
So the combination of both methods leaves little room for doubts.

Moreover this works sheds a new light on the T-system. 
The fact that it should be understood as an operator identity where the product is the fusion of line operators, may be helpful to understand better the integrable structures that appear in the AdS/CFT correspondence.

\paragraph{The role of the pure spinor ghosts.}
The pure spinor ghosts are expected to play an important role in the quantum worldsheet theory.
We can ask the question of the role of the pure spinor ghosts in the computation described in this paper.
Interestingly, the same results would be obtained if we would set the pure spinor ghosts to zero from the start.
The reason is that at tree level, the ghosts form a closed subsector. 
More precisely the OPE of a ghost current with any other current can only produce ghost current. In other words, all the coefficients of the type $C_{g*}^m$ and $C_{\bar g*}^m$ in the current algebra \eqref{Kalgebra} are zero if the index $m$ is not $g$ or $\bar g$.

The fact that the computation does not relies on the pure spinor ghosts can be tracked back to the fact that we only needed the tree-level current algebra to compute the crucial term that produces the shifts in the T-system.
Presumably this is not going to be the case at higher order.
Continuing the computation of \cite{Bedoya:2010av} to get the full current algebra at second order would be interesting.
Already the pure spinor should play an important role. Hopefully they allow for the cancellation of second-order divergences in the transfer matrices, and they also insure that transfer matrices commute up to second order in perturbation theory.

In this paper we used the pure spinor formalism that allows for a covariant quantization. It may also be instructive to reproduce this computation in the Green-Schwarz formulation of \cite{Metsaev:1998it}.

\paragraph{The algebra of transfer matrices.}
The computations we performed allow to address the question of the algebra of transfer matrices for the pure spinor string on $AdS_5 \times S^5$. 
Naively, the fusion of two transfer matrices is rather complicated. It seems from \eqref{summaryFusionTT} that it does not even close on transfer matrices.
However by selecting a particular combination of representations we managed to close the algebra. For these representations, the algebra of transfer matrices is nothing but the T-system.

There might exists a generalization of the T-system that applies for other representations. It would be interesting to further explore this issue.

\paragraph{Generalization to other integrable field theories.}
It would be also interesting to try the approach advocated here in other integrable theories that play a role in the AdS/CFT correspondence. 
Some examples were listed in section \ref{extensions}.
This approach may be more efficient than trying to reproduce the historical steps that were performed for $AdS_5 \times S^5$. 
More generally, developing worldsheet technology for strings in RR background is certainly worthwhile. 
Even if the progress in that direction have been rather slow, the results presented here together with other recent works (see e.g. \cite{Ashok:2009jw}\cite{Vallilo:2011fj}) demonstrate that quantum string theory in some RR backgrounds can be studied with the tools that are currently available.

The results presented here also suggest that the integrable models relevant for the AdS/CFT correspondence may belong to a special family. It is not clear that the interpretation of the T-system advocated here applies straightforwardly to generic integrable field theories. Indeed  generically the transfer matrices have to be renormalized when the dual Coxeter number of the symmetry group is non-zero (see e.g. \cite{Bachas:2004sy}). 
This would complicate a tentative derivation of the T-system from the fusion of transfer matrices.


\section*{Acknowledgments}

The author would like to thank Gleb Arutyunov, Oscar Bedoya, Denis Bernard, Nikolay Gromov, Volodya Kazakov, Marc Magro, Valentina Puletti, Sakura Schafer-Nameki, Joerg Teschner, Jan Troost, Benoit Vicedo, Dmitro Volin and in particular Pedro Vieira for useful discussions and correspondence.
The author is a Postdoctoral researcher of FWO-Vlaanderen.
This research is supported in part by the Belgian Federal Science Policy Office through the Interuniversity Attraction Pole IAP VI/11 and by FWO-Vlaanderen through project G011410N.


\begin{appendix}

\section{Conventions}\label{conventions}

Let ${t_A}$ be a basis of the generators of the Lie superalgebra.
The metric is defined as:
\be \kappa_{AB} = STr(t_A t_B) \ee
where the supertrace $STr$ is a non-degenerate graded-symmetric inner product. 
We define the inverse metric as:
\be \kappa^{AB} \kappa_{AC} = \delta^B_C \ee
The metric and its inverse are graded-symmetric:
\be \kappa_{AB} = (-)^{AB}\kappa_{BA} \ee
where $(-)^{AB}$ is a minus sign if and only if both indices $A$ and $B$ are fermionic.
An element $X$ of the Lie superalgebra is expanded as:
\be J = J^A t_A \ee
We adopt ``NE-SW" conventions for the contraction of indices. 
Indices are raised and lowered with the metric in the following way:
\be X^A = \kappa^{AB}X_B \qquad ; \qquad X_A = X^B \kappa_{BA} \ee

\paragraph{Structure constants.}
The graded commutator for the generators is defined as:
\be  [t_A,t_B\} = t_A t_B - (-)^{AB} t_B t_A \ee
We define the structure constants ${f_{AB}}^C$ as:
\be [t_A,t_B\} = {f_{AB}}^C t_C \ee
The identity $STr(t_A[t_B,t_C\}) = Str([t_A,t_B\}t_C)$
follows from the graded-symmetry of the supertrace. 
It implies for the structure constants:
\be {f_{BC}}^D \kappa_{AD} = {f_{AB}}^D \kappa_{DC} \ee
In agreement with our conventions we define:
\be f_{ABC} = {f_{AB}}^D \kappa_{DC} \ee
and so on. 
The structure constants are graded-antisymmetric in the 1-2 and 2-3 indices\footnote{
With different conventions (SE-NW), it would be the 1-2 and 1-3 indices.}:
\be f_{ABC} = -(-)^{AB}f_{BAC} \qquad ; \qquad f_{ABC} = -(-)^{BC} f_{ACB} \ee
Under the exchange of the first and third indices, we have:
\be f_{ABC} = -(-)^{\frac{A+B+C}{2}}f_{CBA} \ee
Under cyclic permutation of their indices, the structure constant also satisfy:
\be f_{ABC} = (-)^A f_{BCA}  
\ee

\paragraph{Tensor product.}
The tensor product $t_A^R \otimes t_B^{R'}$ of two generators taken in different representation $R$ and $R'$ is graded:
\be t_A^R \otimes t_B^{R'} = (-)^{AB} (1^R \otimes t_B^{R'})(t_A^R \otimes 1^{R'}) \ee
In order to lighten the expressions in the bulk of the paper we often get rid of the $\otimes$ symbol:
\be t_A^R \otimes t_B^{R'} \equiv t_A^R  t_B^{R'}  = (-)^{AB} t_B^{R'} t_A^R \ee


\section{A new look at the gauge covariant current algebra}\label{covCurrents}

\subsection{Derivation of the current-current OPEs}

In this appendix we give a new derivation of the tree-level gauge-covariant current algebra for the pure spinor string on $AdS_5 \times S^5$. The method we use is inspired by the analysis of \cite{Benichou:2010rk} for the current algebra in sigma-models on supergroups. 
The different steps are the following. We make a natural ansatz for the current-current OPEs. Then we demand that this ansatz is compatible with reparametrization invariance of the path integral, and the with Maurer-Cartan equation. This typically gives more constraints than the number of free coefficients in the ansatz. Finally we solve these constraints to get the current algebra. 

This method can be generalized to compute the quantum corrections to the current algebra. This computation can be efficiently organized recursively \cite{Benichou:2010rk}. Here we will only compute the tree-level coefficients since this is sufficient for the purpose of this article.

We choose the ansatz \eqref{Kalgebra} for the current algebra, that we reproduce here for clarity:
\begin{align} \label{ansatzKK}
K_m^{A_m}(z) K_n^{B_n}(w)  = &
R^{-2} C_{mn} \frac{\kappa^{B_n A_m}}{(z-w)^2}
+ R^{-2} \sum_p C_{mn}^p \frac{{f_{C_p}}^{B_n A_m} K_p^{C_p}}{z-w} \cr
& + R^{-2} \sum_p C_{mn}^{\bar p} {f_{C_p}}^{B_n A_m} \bar K_p^{C_p} \frac{\bar z - \bar w}{(z-w)^2} +... \cr
K_m^{A_m}(z) \bar K_n^{B_n}(w)  = &
R^{-2} C_{m \bar n} \kappa^{B_n A_m} 2\pi \delta^{(2)}(z-w)
+ R^{-2} \sum_p C_{m\bar n}^p \frac{{f_{C_p}}^{B_n A_m} K_p^{C_p}}{\bar z-\bar w} \cr
& + R^{-2} \sum_p C_{m\bar n}^{\bar p} \frac{{f_{C_p}}^{B_n A_m} \bar K_p^{C_p}}{z-w} +...\cr
\bar K_m^{A_m}(z) \bar K_n^{B_n}(w)  = &
R^{-2} C_{\bar m \bar n} \frac{\kappa^{B_n A_m} }{(\bar z - \bar w)^2}
+ R^{-2} \sum_p C_{\bar m\bar n}^p {f_{C_p}}^{B_n A_m} K_p^{C_p}\frac{z-w}{(\bar z-\bar w)^2} \cr
& + R^{-2} \sum_p C_{\bar m\bar n}^{\bar p} \frac{{f_{C_p}}^{B_n A_m} \bar K_p^{C_p}}{\bar z-\bar w} +...
\end{align}
This ansatz is based on dimensional analysis and symmetry. We only wrote down the second- and first-order poles, but there is an infinite series of less and less singular terms that come with operators of (classical) dimension greater or equal to two. Notice also that this ansatz is suitable for the tree-level current algebra, but it should be slightly modified if one is to take into account quantum corrections \cite{Bedoya:2010av}.
In the following we will compute the coefficients $C$'s.
Many of these coefficient vanish trivially because the current algebra has to be compatible with the $\mathbb{Z}_4$ grading.
Parity also induces some redundancy in the remaining coefficients.
These two symmetries leave 57 independent coefficients that we need to compute.


\paragraph{Equations of motion and path-integral reparametrization invariance.}

In this subsection we demand that the current algebra \eqref{ansatzKK} is compatible with the reparametrization invariance of the path integral. In particular this guarantees that the current algebra is compatible with the equations of motion.

Let us consider the action for the pure spinor string in $AdS_5 \times S^5$ \eqref{action}. 
We consider a small variation of the group element $g$ parametrized by a element of the Lie superalgebra $X$:
\be\label{shiftG} \delta g = g X \ee%
The variation of the currents is given by:
\begin{align}  & \delta J = \p X + [J,X] && \delta N = 0\cr
&\delta \bar J = \bar \p X + [\bar J,X]  & &\delta \hat N = 0 \end{align}
We can decompose the infinitesimal shift $X$ on the $\mathbb{Z}_4$ subspaces of the Lie superalgebra as $X=X_0+X_1+X_2+X_3$. Since $X_0$ generates gauge transformations that leave the action invariant, we set $X_0 = 0$. We obtain the variation of the $J_i$'s:
\beq\label{deltaJ} \delta J_{0} &=&  [J_{1},X_3] + [J_{2},X_2] + [J_{3},X_3] \cr
\delta J_{1} &=&  \p X_1 + [J_{0},X_1] + [J_{2},X_3] + [J_{3},X_2] \cr
\delta J_{2} &=&  \p X_2 + [J_{0},X_2] + [J_{1},X_1] + [J_{3},X_3] \cr
\delta J_{3} &=&  \p X_3 + [J_{0},X_3] + [J_{1},X_2] + [J_{2},X_1] \eeq
and similarly for the $\bar J_i$'s. 
We deduce the variation of the action under the infinitesimal shift of the group element \eqref{shiftG}:
\begin{align}\label{deltaS} \delta  S =   \frac{R^2}{4\pi} STr \int d^2 z & \left\{
 X_1  \left( - \frac{3}{2} \bar \nabla J_{3} - \frac{1}{2} \nabla \bar J_{ 3} - \frac{1}{2} [J_{1},\bar J_{2}]- \frac{1}{2} [J_{2}, \bar J_{1}] + 2[N,\bar J_3] - 2 [J_3, \hat N] \right) \right.\cr
& +   X_2 \left(- \bar \nabla J_{2} -  \nabla \bar J_{ 2} -  [J_{1},\bar J_{1}]+  [J_{3},\bar J_{3}] + 2[N,\bar J_2] - 2 [J_2, \hat N] \right) \cr
& \left.+  X_3 \left( -\frac{1}{2} \bar \nabla J_{1} - \frac{3}{2} \nabla \bar J_{ 1} + \frac{1}{2} [J_{2},\bar J_{3}]+ \frac{1}{2} [J_{3},\bar J_{2}]+ 2[N,\bar J_1] - 2 [J_1, \hat N]\right) \right\}
 \end{align}
Now we consider the following quantity:
\be \langle J_1(z) \rangle = \int \mathcal{D}\Phi J_1(z) e^{-S} \ee
where $ \mathcal{D}\Phi $ is the path integral measure over the fields. The previous one-point function, whatever its value is, is invariant under the reparametrization of the path integral \eqref{shiftG}. We further assume that the path-integral measure is also invariant under \eqref{shiftG}.
Let us mention at that point that we are simply following the method that would provide a path integral derivation of the Ward identity for a global symmetry, if \eqref{shiftG} were indeed a global symmetry.
We obtain:
\be\label{dJ=JdS} \langle \delta J_1(z) - J_1(z) \delta S \rangle = 0 \ee
It is convenient to rewrite the variation of the current \eqref{deltaJ} as an integral over the worldsheet:
\be \delta J_1(z) = \int d^2 w \left( X_1(w) \delta'(z-w) + \left([J_0(w),X_1(w)]+[J_2(w),X_3(w)]+[J_3(w),X_2(w)]\right) \delta(z-w) \right) \ee
Projecting equation \eqref{dJ=JdS} on the $\mathbb{Z}_4$ subspaces, we obtain three operator identities:
\begin{align}\label{appJ1EOM}  J_1^{A_1}(z) & \left(-\frac{3}{2} \bar \nabla J_3^{B_3}(w) - \frac{1}{2} \nabla \bar J_3^{B_3}(w) - \frac{1}{2}{f_{C_1 D_2}}^{B_3} :J_1^{C_1} \bar J_2^{D_2}:(w) - \frac{1}{2} {f_{C_2 D_1}}^{B_3} :J_2^{C_2} \bar J_1^{D_1}:(w) \right. \cr
& \left. \qquad + 2 {f_{C_0 D_3}}^{B_3} :N^{C_0} \bar J_3^{D_3}:(w) - 2 {f_{C_3 D_0}}^{B_3} :J_3^{C_3} \hat N^{D_0}:(w) \right) \cr
& = 4 \pi R^{-2} \kappa^{B_3 A_1} \p_z \delta(z-w) \cr
J_1^{A_1}(z) & \left(- \bar \nabla J_2^{B_2}(w) - \nabla \bar J_2^{B_2}(w) - {f_{C_1 D_1}}^{B_2} :J_1^{C_1} \bar J_1^{D_1}:(w) + {f_{C_3 D_3}}^{B_2} :J_3^{C_3} \bar J_3^{D_3}:(w) \right. \cr
& \left.\qquad  + 2 {f_{C_0 D_2}}^{B_2} :N^{C_0} \bar J_2^{D_2}:(w) - 2 {f_{C_2 D_0}}^{B_2} :J_2^{C_2} \hat N^{D_0}:(w) \right) \cr
&  = 4 \pi R^{-2}  {f_{C_3}}^{B_2 A_1} J_3^{C_3}(w) \delta(z-w) \cr 
J_1^{A_1}(z) & \left(-\frac{1}{2} \bar \nabla J_1^{B_1}(w) - \frac{3}{2} \nabla \bar J_1^{B_1}(w) + \frac{1}{2}{f_{C_2 D_3}}^{B_1} :J_2^{C_2} \bar J_3^{D_3}:(w) + \frac{1}{2} {f_{C_3 D_2}}^{B_1} :J_3^{C_3} \bar J_2^{D_2}:(w) \right. \cr
& \left. \qquad + 2 {f_{C_0 D_1}}^{B_1} :N^{C_0} \bar J_1^{D_1}:(w) - 2 {f_{C_1 D_0}}^{B_1} :J_1^{C_1} \hat N^{D_0}:(w) \right) \cr
&  = 4 \pi R^{-2} {f_{C_2}}^{B_1 A_1} J_2^{C_2}(w) \delta(z-w)
\end{align}
where the colons stand for normal ordering.
Next we plug the ansatz \eqref{ansatzKK} into these equations. More precisely, we use the ansatz \eqref{ansatzKK} to perform the OPEs on the left-hand side of the identities \eqref{appJ1EOM}. Since we are working at first-order in $R^{-2}$, the OPEs involving composite operators are easily dealt with: a single OPE has to be taken with one or the other of the components of the composite operator.
 We use the equalities:
\be \p_{\bar w} \frac{1}{z-w} = -2\pi \delta(z-w) = \p_w \frac{1}{\bar z - \bar w} 
\quad ; \quad 
 \p_{\bar w} \frac{1}{(z-w)^2} = 2\pi \delta'(z-w) = \p_w \frac{1}{(\bar z - \bar w)^2}\ee
We are left with some identities between operators multiplied by functions that are singular when $z-w \to 0$.
We demand that the operator identities \eqref{appJ1EOM} do hold for the singular terms of order two and three: all the terms multiplying either a derivative of a delta function, a delta function, or a second-order pole shall cancel against each other\footnote{Demanding that the singular terms of order one or less do also vanish is not consistent with the ansatz \eqref{ansatzKK}, since the subleading terms in the current algebra that we did not write in \eqref{ansatzKK} would contribute \cite{Benichou:2010rk}.}.  We obtain a set of linear equations that the free coefficients in the ansatz \eqref{ansatzKK} have to satisfy:
\begin{align}
& 2 = -\frac{3}{2} C_{13} + \frac{1}{2} C_{1 \bar 3}
& 0 = \frac{3}{2} C_{13}^{\bar g} - \frac{1}{2} C_{1 \bar 3}^{\bar g} - 2 C_{13} 
& \qquad 0 = \frac{3}{2} C_{13}^g + \frac{1}{2} C_{1 \bar 3}^g - 2C_{1 \bar 3} \cr
& 0 = \frac{3}{2} C_{13}^{\bar 0} - \frac{1}{2} C_{1 \bar 3}^{\bar 0} + \frac{3}{2} C_{13}
& 2 = \frac{3}{2} C_{13}^0 + \frac{1}{2} C_{1 \bar 3}^0 + \frac{1}{2} C_{1 \bar 3}
& \qquad 0 = C_{12}^{\bar 3} - C_{1 \bar 2}^{\bar 3} + C_{13} \cr
& 2 = C_{12}^3 + C_{1 \bar 2}^3 - C_{1 \bar 3} 
& 0 = \frac{1}{2} C_{11}^{\bar 2} - \frac{3}{2} C_{1 \bar 1}^{\bar 2} + \frac{1}{2} C_{13}
& \qquad 2 = \frac{1}{2} C_{11}^2 + \frac{3}{2} C_{1 \bar 1}^{2} - \frac{1}{2} C_{1 \bar 3}
\end{align}
We can play the same game replacing in equation \eqref{dJ=JdS} $J_1$ by another current. For each current we obtain a new set of equations. 
To get more constraints for the OPEs involving the ghosts, we can also vary the ghosts variable instead of \eqref{shiftG}.
In total we get 39 linear equations that the coefficients $C$'s in the ansatz \eqref{ansatzKK} have to satisfy.


\paragraph{The Maurer-Cartan equation.}

We can further constraint the coefficients in the ansatz \eqref{ansatzKK} by demanding compatibility with the Maurer-Cartan equation \eqref{MC}.
Projecting this equation according to the $\mathbb{Z}_4$ grading we obtain:
\begin{align} \label{MCcomp}
& \p \bar J_0 - \bar \p J_0 + [J_0, \bar J_0] + [J_1, \bar J_3] + [J_2, \bar J_2] + [J_3, \bar J_1] = 0 \cr
& \p \bar J_1- \bar \p J_1 + [J_0, \bar J_1] + [J_1, \bar J_0] + [J_2, \bar J_3] + [J_3, \bar J_2] = 0 \cr
& \p \bar J_2 - \bar \p J_2 + [J_0, \bar J_2] + [J_1, \bar J_1] + [J_2, \bar J_0] + [J_3, \bar J_3] = 0 \cr
& \p \bar J_3 - \bar \p J_3 + [J_0, \bar J_3] + [J_1, \bar J_2] + [J_2, \bar J_1] + [J_3, \bar J_0] = 0 
 \end{align}
The strategy is to demand that the Maurer-Cartan equation does hold as an operator identity. More precisely, we demand that the OPE between a current and the left-hand side of \eqref{MC} does vanish. Let us consider one example for illustrative purposes: we take the OPE between the current $J_1$ and the left-hand side of the first line in equation \eqref{MCcomp}:
\begin{align}
0 = J_1^{A_1}(z) & \left( \p \bar J_0^{B_0}(w) - \bar\p J_0^{B_0}(w) + {f_{C_0 D_0}}^{B_0} :J_0^{C_0} \bar J_0^{D_0}:(w) + {f_{C_1 D_3}}^{B_0} :J_1^{C_1} \bar J_3^{D_3}:(w) \right. \cr
& \left. \qquad +  {f_{C_2 D_2}}^{B_0} :J_2^{C_2} \bar J_2^{D_2}:(w) + {f_{C_3 D_1}}^{B_0} :J_3^{C_3} \bar J_1^{D_1}:(w) \right) 
\end{align}
As previously we plug the ansatz \eqref{ansatzKK} in the previous equation, and demand that the singular terms of order three and two do vanish. We obtain the following equations:
\be 
0 = C_{10}^{\bar 1} + C_{1 \bar 0}^{\bar 1} + C_{13}
\qquad 
0 = C_{10}^1 - C_{1 \bar 0}^1 - C_{1 \bar 3}
\ee
Repeating the same procedure for the different current and the different lines of equation \eqref{MCcomp}, we obtain in total 43 linear equations that the coefficients $C$'s in the ansatz \eqref{ansatzKK} have to satisfy.

Let us make a side remark here. There is no doubt that the Maurer-Cartan identity does hold at tree level. However it may get quantum corrections. In order to generalize the method described here to compute quantum corrections to the current algebra, one needs to assume that the Maurer-Cartan identity holds in the quantum theory as well. This may be interpreted as postulating quantum integrability of the model. This provides an efficient way to use quantum integrability of the model to compute the quantum current algebra. We leave it for future work.


\paragraph{The coefficients of the current algebra.}

Using the Maurer-Cartan identity and reparametrization invariance of the path integral, we find in total 82 equations that constrain the 57 independent coefficients of the current algebra \eqref{ansatzKK}. 
This system of equation can be easily decomposed in subsystems of eight equations or less.
It is remarkable that there exists a solution to this set of equations. The non-zero coefficients are given in section \ref{subKalgebra}.

There is however one exception for the OPEs between two of the currents $J_0$ and $\bar J_0$. In that case the equations we obtain only provide the constraints \eqref{constraintsC00*}.


\paragraph{Associativity.}
The current algebra has to be associative. Associativity of the current algebra can be tested in the following way. Let us consider a 3-points function, for instance:
\be \langle J_1(x) J_1(y) J_2(z) \rangle \ee
It can be computed by taking first the OPE between $J_1(x)$ and $ J_1(y)$, and then take the OPE between the resulting current and $J_2(z)$. 
But one can also start by taking the OPE between $J_1(y)$ and $J_2(z)$, and then take the OPE of the result with $J_1(x)$. The two methods lead to the same result if the coefficients of the current algebra satisfy:
\be C_{11}^2 C_{22} = C_{12}^3 C_{13} \ee
We can play the same game with any three-points functions. We find a large set of constraints that are all satisfied by the current algebra obtained previously.


\subsection{The $(r,s)$ system.}\label{appRSmatrices}
In this section we give some details on the derivations of equations \eqref{r,sSyst}, \eqref{rMatrix} and \eqref{sMatrix}.
We want to compute the commutator of two equal-time connections $A_R(y;\sigma)$ and $A_{R'}(y';\sigma')$ evaluated for different values of the spectral parameter $y$ and $y'$ and taken in possibly different representation $R$ and $R'$. We can deduce this commutator from the current algebra. 
We define the commutator of equal-time operators as:
\be [A(\sigma),B(0)] = \lim_{\epsilon \to 0^+}  \left( A(\sigma+i \epsilon)B(0) - B(i \epsilon)A(\sigma) \right) \ee
From this definition we extract an operative dictionary between OPEs and commutators.
Let us consider for instance the following OPE:
\be\label{ABOPE} A(z) B(0) = \frac{C}{z^2} + \frac{D}{\bar z^2} + E \delta^{(2)}(z) + \frac{F(0)}{z} + \frac{G(0) \bar z}{z^2} + \frac{H(0)}{\bar z} + \frac{I(0) z}{\bar z^2} + ... \ee
We deduce the commutator:
\be\label{ABcom} \frac{1}{2\pi i} [ A(\sigma), B(0) ] = C \delta'(\sigma) - D \delta'(\sigma) - F(0) \delta(\sigma) - G(0)\delta(\sigma) + H(0)\delta(\sigma) + I(0)\delta(\sigma) \ee
This dictionary shows that the first-order computation of fusion presented in section \ref{fusionOrder1} is equivalent to the computation of the Poisson bracket of line operators in the Hamiltonian formalism. 
Notice that the OPE \eqref{ABOPE} generically contains additional sub-leading singularities, for instance $\frac{\bar z}{z}$, or even logarithms. They do not contribute to the commutator \cite{Benichou:2010ts}. Generically, the OPE contains more information than the commutator.

In order to simplify the following expressions, we write the (spacelike component of the) flat connection $A(y)$ defined in \eqref{defA} as:
\begin{align}\label{defAwithFs} A(y) =& \sum_m F_m(y) K_m + \bar F_m(y) \bar K_m \end{align}
Using the previous dictionary, we obtain for the commutator of two connections:
\begin{align}\label{[A,A']}
 [A_R&(y;\sigma),  A_{R'}(y';\sigma')]  = \cr
& 2\pi i R^{-2} \p_\sigma \delta(\sigma- \sigma') \sum_{m,n} \kappa^{B_n A_m} t_{A_m}^R t_{B_n}^{R'} ( F_m(y)F_n(y')C_{mn} - 
 \bar F_m(y)\bar F_n(y')C_{\bar m \bar n}) \cr
& + 2\pi i R^{-2}  \delta(\sigma- \sigma') \sum_{m,n,p} {f_{C_p}}^{B_n A_m} t_{A_m}^R t_{B_n}^{R'}
K_p^{C_p} ( -F_m(y)F_n(y') C_{mn}^p \cr
& \hspace{2.5 cm} + F_m(y)\bar F_n(y') C_{m\bar n}^p + \bar F_m(y)F_n(y') C_{\bar m n}^p + \bar F_m(y) \bar F_n(y') C_{\bar m \bar n}^p) \cr
& + 2\pi i R^{-2}  \delta(\sigma- \sigma') \sum_{m,n,p} {f_{C_p}}^{B_n A_m} t_{A_m}^R t_{B_n}^{R'} 
\bar K_p^{C_p} ( -F_m(y)F_n(y') C_{mn}^{\bar p}  \cr
& \hspace{2.5 cm} - F_m(y)\bar F_n(y') C_{m\bar n}^{\bar p} - \bar F_m(y)F_n(y') C_{\bar m n}^{\bar p} + \bar F_m(y) \bar F_n(y') C_{\bar m \bar n}^{\bar p}) 
\end{align}
We wish to write this commutator as a $(r,s)$ system \eqref{r,sSyst}. From the terms coming with a derivative of the delta function in the commutator \eqref{[A,A']}, we can read directly the $s$-matrix. We obtain:
\be s = \pi i R^{-2} \sum_{m,n} \kappa^{B_n A_m} t_{A_m}^R t_{B_n}^{R'} \left( F_m(y)F_n(y')C_{mn} - 
 \bar F_m(y)\bar F_n(y')C_{\bar m \bar n}\right) \ee
Plugging in the value of the coefficients, we obtain \eqref{sMatrix}. To obtain the $r$ matrix, we have to compare the terms coming with a delta function in \eqref{[A,A']} and \eqref{r,sSyst}. This leads to the following equations for the components of the $r$ and $s$ matrices:
\begin{align}\label{eqsForRMatrix} \forall\ m,n,p: \quad 
F_p(y) & r_{4-n,n} - F_p(y') r_{m,4-n} = - F_p(y) s_{4-n,n} - F_p(y') s_{m,4-n}  - 2 F_m(y) F_n(y') C_{m n}^{p} \cr
&+ 2 F_m(y) \bar F_n(y') C_{m \bar n}^{p}+ 2 \bar F_m(y) F_n(y') C_{\bar m n}^{p} + 2 \bar F_m(y) \bar F_n(y') C_{\bar m \bar n}^{p} \cr \cr
\bar F_p(y) & r_{4-n,n} - \bar F_p(y') r_{m,4-n} = - \bar F_p(y) s_{4-n,n} - \bar F_p(y') s_{m,4-n}  - 2 F_m(y) F_n(y') C_{m n}^{\bar p} \cr
&- 2 F_m(y) \bar F_n(y') C_{m \bar n}^{\bar p}- 2 \bar F_m(y) F_n(y') C_{\bar m n}^{\bar p} + 2 \bar F_m(y) \bar F_n(y') C_{\bar m \bar n}^{\bar p}
\end{align}
In the previous equations, when the indices $m,n$ take the value $0$ or $g$, one should understand``$r_{4,0}$" and ``$r_{4-g,g}$" as being $r_{0,0}$,  etc.
Remarkably, this largely over-constrained system is solved by the $r$ and $s$ matrices \eqref{rMatrix} and \eqref{sMatrix}.


\paragraph{Comparison with previous analyses.}\label{appCompKAlgebras}
The current-current OPEs were previously discussed in the literature. 
In \cite{Puletti:2006vb} the OPEs for the currents of non-zero grade were computed using the background field methods. Some of the OPEs involving the grade zero currents were further given in \cite{Puletti:2008ym}. The results we obtained here agree with these papers.

In \cite{Mikhailov:2007mr} the current algebra was also computed using Feynman diagram technology.
The OPEs do match the ones we derived here except for those involving the currents $J_0$, $\bar J_0$. This is not surprising given the gauge choice that was explicitly made for the coset element in \cite{Mikhailov:2007mr}.
A consequence of this discrepancy is that the commutator of equal-time connections can not be written as a $(r,s)$ system with the OPEs of \cite{Mikhailov:2007mr}. However one should keep in mind that it is only an issue of gauge fixing.
Indeed in \cite{Mikhailov:2007eg} the OPEs of \cite{Mikhailov:2007mr} were used to compute the fusion of line operators at first order. Then the $r$ and $s$ matrices were deduced by comparison with the expectations from the Hamiltonian formalism. These matrices agree with the ones that we derived in this paper.

In \cite{Magro:2008dv} the hamiltonian formalism was used to compute the commutator of equal time connections. A careful treatment of the constraints was performed. 
It was argued that in the Hamiltonian formalism, the flat connection \eqref{defA} realizes a $(r,s)$ system up to constraints generating gauge transformations.
It was shown that one should add to the flat connection a term proportional to the constraints so that the commutator of connections take exactly the form of a $(r,s)$ system. 
The resulting $(r,s)$ system is slightly different than the one used here and in \cite{Mikhailov:2007mr}. 
The flat connection obtained in \cite{Magro:2008dv}, including the additional term proportional to the constraints, was derived from first principles in \cite{Vicedo:2009sn} in the Green-Schwarz formalism.
It is remarkable that the analysis of \cite{Vicedo:2009sn} leads to the pure spinor-like flat connections of \cite{Vallilo:2003nx} (without the pure spinor ghosts contribution) and not to the Bena-Polchinski-Roiban flat connections \cite{Bena:2003wd}.
This provides some evidence for the equivalence of the pure spinor and Green-Schwarz formulations of string theory on $AdS_5 \times S^5$.
In \cite{Vicedo:2010qd} it was shown that the $(r,s)$ system of \cite{Magro:2008dv} has a nice algebraic interpretation.

For the purposes of this paper it is important that the $r$ matrix found in \cite{Magro:2008dv} is identical to the one we worked with in the limit where the difference of spectral parameter is small \eqref{rLim}.
This guarantees that the results derived in the present paper would also hold if one were to work with the $(r,s)$ system of \cite{Magro:2008dv}.
In order to reproduce the $(r,s)$ system found in \cite{Magro:2008dv} using OPEs technology, the first step would be to gauge-fix the $\mathcal{H}_0$ gauge symmetry via a BRST procedure. Then one should generalize the analysis of \cite{Vallilo:2003nx} by including in the flat connections additional terms written in terms of the ghosts resulting from the $\mathcal{H}_0$ gauge-fixing. These new connections should realize the $(r,s)$ system of \cite{Magro:2008dv}\footnote{The author would like to thank B. Vicedo for illuminating discussions on this point.}.

%
%


\section{Divergences in line operators}\label{computationDiv}

In this appendix we give some details about the computations of the first-order divergences in line operators. 
As explained in section \ref{worldsheetTheory}, we use a principal-value regularization scheme. A first-order pole is regularized as:
\be \frac{1}{\sigma-\sigma'} \to P.V.\frac{1}{\sigma-\sigma'} = \frac{1}{2}\left(\frac{1}{\sigma+i\epsilon-\sigma'} + \frac{1}{\sigma-i\epsilon-\sigma'}\right) = \frac{\sigma-\sigma'}{(\sigma-\sigma')^2 + \epsilon^2} \ee
and a second-order pole is regularized as:
\be \frac{1}{(\sigma-\sigma')^2} \to P.V.\frac{1}{(\sigma-\sigma')^2} = \frac{1}{2}\left(\frac{1}{(\sigma+i\epsilon-\sigma')^2} + \frac{1}{(\sigma-i\epsilon-\sigma')^2}\right) = \frac{(\sigma-\sigma')^2-\epsilon^2}{((\sigma-\sigma')^2 + \epsilon^2)^2} \ee

\subsection{Divergences in transition matrices}

\begin{figure}
\centering
\includegraphics[scale=0.50]{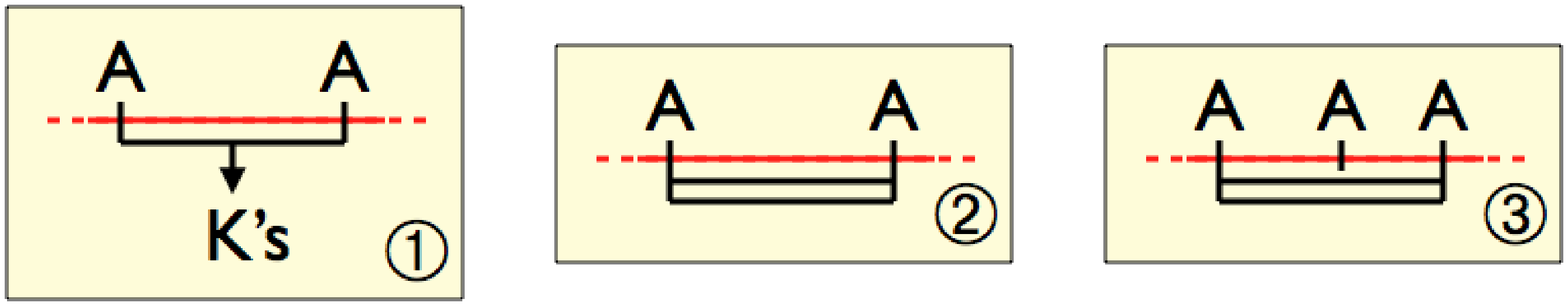}
\caption{The three individual sources of first-order divergences in line operators. In \ding{172} the first-order poles in the OPE between two connections are considered. In \ding{173} and \ding{174} the second-order poles are considered.\label{divergences}}
\end{figure}

There are three sources of divergences in the transition matrices.
They are depicted in figure \ref{divergences}.
The first divergences come from the first order poles in the OPE of two neighboring connections, say $A(y;\sigma_1)$ and $A(y;\sigma_2)$ (case \ding{172} in figure{divergences}). We evaluate the resulting currents at the point $\sigma_2$ and perform the integration over $\sigma_1$. We obtain a logarithmic divergences:
\begin{align}\label{divAA1}
& (-\log \epsilon) \sum_{m,n,p} \left( 
K_p^{C_p}(\sigma_2)(F_m F_n C_{mn}^p + \bar F_m F_n C_{\bar m n}^p + F_m \bar F_n C_{m \bar n}^p + \bar F_m \bar F_n C_{\bar m \bar n}^p) \right. \cr
& \left. + \bar K_p^{C_p}(\sigma_2)(F_m F_n C_{mn}^{\bar p} + \bar F_m F_n C_{\bar m n}^{\bar p} + F_m \bar F_n C_{m \bar n}^{\bar p} + \bar F_m \bar F_n C_{\bar m \bar n}^{\bar p}) \right)
{f_{C_p}}^{B_n A_m} t_{A_m} t_{B_n}
\end{align}
where the functions $F$'s were defined in \eqref{defAwithFs}.

The second type of divergences come from the second-order poles in the OPE between two neighboring connections (case \ding{173} in figure{divergences}). After performing the integration over the positions of the two connections, we obtain a logarithmic divergence:
\begin{align}\label{divAA2}
& \log \epsilon \sum_{m,n} 
(F_m F_n C_{mn} + \bar F_m \bar F_n C_{\bar m \bar n}) \kappa^{B_n A_m} t_{A_m} t_{B_n}
\end{align}
Notice that there is no linear divergences. This is a pleasant feature of the regularization scheme that we are using.
Eventually the third type of divergences come from the second-order poles in the OPE between two connections that are separated by a third one sitting in between (case \ding{174} in figure{divergences}). Let us denote this third connection by $A(y;\sigma)$. After performing the integrations, we obtain another logarithmic divergence:
\begin{align}\label{divAAA2}
& (-\log \epsilon) \sum_{m,n} 
 (F_m F_n C_{mn} + \bar F_m \bar F_n C_{\bar m \bar n}) 
\left( \frac{1}{2}\left\{ \kappa^{B_n A_m} t_{A_m} t_{B_n} ,A(y;\sigma) \right\} \right. \cr
& \quad \left. -\frac{1}{2} \sum_{p,q} \left({f_{C_p}}^{D_q A_m} t_{A_m} t_{D_q} + {f_{C_p}}^{B_n D_q}t_{D_q} t_{B_n}\right)\left(F_p K_p^{C_p}(\sigma) + \bar F_p \bar K_p^{C_p}(\sigma)\right) \right)
\end{align}
There is some freedom in how we write the last expression since we can commute the generators in different ways. We choose a writing that is symmetric with respect to the central connection $A(y;\sigma)$.

Starting from a transition matrix, we compute all the different OPEs of the types described previously that lead to divergences.
Next we sum all these terms. 
Most of the terms of the type \eqref{divAA2} cancel against the first terms in \eqref{divAAA2}. We obtain:
\begin{align}\label{divTot1}
\sum_{M=0}^\infty&\left((-)^M \log \epsilon \frac{1}{2} \sum_{m,n} (F_m F_n C_{mn} + \bar F_m \bar F_n C_{\bar m \bar n}) 
\left\{ \kappa^{B_n A_m} t_{A_m} t_{B_n}, \left \lfloor \int_a^b A \right \rceil^M \right\}\right.\cr
& + (-)^{M+1} \log \epsilon \sum_{i=0}^{M-1} \int_a^b d \sigma \left \lfloor \int_\sigma^b A \right \rceil^i
\sum_{m,n,p} {f_{C_p}}^{B_n A_m} t_{A_m} t_{B_n}  \cr
& \qquad \times ( K_p^{C_p}(\sigma) (F_m F_n C_{mn}^p + \bar F_m F_n C_{\bar m n}^p + F_m \bar F_n C_{m \bar n}^p + \bar F_m \bar F_n C_{\bar m \bar n}^p  \cr
& \qquad \qquad +\frac{1}{2}F_p \sum_q(F_m F_q C_{mq} + \bar F_m \bar F_q C_{\bar m \bar q} + F_n F_q C_{nq} + \bar F_n \bar F_q C_{\bar n \bar q})) \cr
& \qquad \quad  + \bar K_p^{C_p}(\sigma)(F_m F_n C_{mn}^{\bar p} + \bar F_m F_n C_{\bar m n}^{\bar p} + F_m \bar F_n C_{m \bar n}^{\bar p} + \bar F_m \bar F_n C_{\bar m \bar n}^{\bar p} \\
& \left.\qquad \qquad +\frac{1}{2}\bar F_p \sum_q(F_m F_q C_{mq} + \bar F_m \bar F_q C_{\bar m \bar q} + F_n F_q C_{nq} + \bar F_n \bar F_q C_{\bar n \bar q})))
\left \lfloor \int_a^\sigma A \right \rceil^{M-i-1} \right) \nonumber
\end{align}

\paragraph{Consequences of the vanishing of the dual Coxeter number.}
Here we derive some identities that are useful to show the vanishing of some divergences in the line operators. These identities were first derived in \cite{Mikhailov:2007mr}.
The vanishing of the dual Coxeter number can be written as:
\be\label{ftt=0}  {f_C}^{BA}[t_A,t_B\} = 0 \ee
The super-Jacobi identity together with the fact that $\kappa^{A_1 B_3} \{t_{A_1},t_{B_3}\} = 0$ implies:
\be\label{13C=31C} \kappa^{B_3 A_1}[t_{A_1},[t_{B_3},t_C\}\} = \kappa^{A_1 B_3}[t_{B_3},[t_{A_1},t_C\}\} \ee
The identities \eqref{ftt=0} and \eqref{13C=31C} further imply:
\begin{align}\label{C=1}
{f_{C_1}}^{D_2 B_3}[t_{B_3},t_{D_2}\} = {f_{C_1}}^{D_0 A_1}[t_{A_1},t_{D_0}\} = 0 \end{align}
\begin{align}\label{C=3}
{f_{C_3}}^{D_2 B_1}[t_{B_1},t_{D_2}\} = {f_{C_3}}^{D_0 A_3}[t_{A_3},t_{D_0}\} = 0 \end{align} 
\begin{align}\label{C=2}
{f_{C_2}}^{D_1 A_1}[t_{A_1},t_{D_1}\} = {f_{C_2}}^{D_3 A_3}[t_{A_3},t_{D_3}\} = - {f_{C_2}}^{D_0 B_2}[t_{B_2},t_{D_0}\} \end{align}

\paragraph{Cancellation of divergences.}
Let us come back to the expression \eqref{divTot1}. We will now argue that the second piece of \eqref{divTot1} vanishes, as first shown in \cite{Mikhailov:2007mr}. Using the identities \eqref{C=1} and \eqref{C=3}, we observe that the terms proportional to $J_1$, $\bar J_1$, $J_3$ and $\bar J_3$ vanish straight away. Then using the identities \eqref{C=2} together with the actual value of the coefficients of the current algebra, it is straightforward to check that the terms proportional to $J_2$ and $\bar J_2$ also drop out. 

The vanishing of the terms proportional to $J_0$, $\bar J_0$, $N$ and $\hat N$ depends on the value of the simple poles in the OPEs $J_0.J_0$, $J_0. \bar J_0$ and $\bar J_0 . \bar J_0$. The method explained in appendix \ref{covCurrents} to compute the current algebra does not fix completely these OPEs, but only gives the constraint \eqref{constraintsC00*}.
The identity \eqref{ftt=0} combined with the value of the other coefficients of the current algebra implies the vanishing of all terms proportional to $J_0$, $\bar J_0$, $N$ and $\hat N$ provided we have:
\begin{align}\label{possibleValueC00*}
C_{00}^0 = C_{0\bar 0}^0 = -C_{\bar 0 \bar 0}^0 = 0 \quad ; \quad 
- C_{00}^{\bar 0} = C_{0\bar 0}^{\bar 0} = C_{\bar 0 \bar 0}^{\bar 0} = 0\cr
C_{00}^g = C_{0\bar 0}^g = -C_{\bar 0 \bar 0}^g = 2 \quad ; \quad 
- C_{00}^{\bar g} = C_{0\bar 0}^{\bar g} = C_{\bar 0 \bar 0}^{\bar g} = 2
\end{align}
Demanding consistency with the analysis of \cite{Mikhailov:2007mr} implies the previous equations.
We deduce that only the first term in \eqref{divTot1} survives. We can write it as:
\begin{align}
&(-)^{M+1} \log \epsilon \frac{y^4 + y^{-4}}{2}  
\left\{ \kappa^{B_3 A_1} t_{A_1} t_{B_3} + \kappa^{B_2 A_2} t_{A_2} t_{B_2} + \kappa^{B_1 A_3} t_{A_3} t_{B_1}, \left \lfloor \int_a^b A \right \rceil^M \right\}\cr
\end{align}
So the first-order divergences in the transition matrix can be rewritten as:
\be\label{divTransition} -\log \epsilon \frac{y^4 + y^{-4}}{2}  \left\{ \kappa^{B_3 A_1} t_{A_1} t_{B_3} + \kappa^{B_2 A_2} t_{A_2} t_{B_2} + \kappa^{B_1 A_3} t_{A_3} t_{B_1}, T^{b,a}(x) \right\} \ee


\subsection{Divergences in monodromy and transfer matrices}

In loop operators we have additional divergences coming from the collisions between two connections sitting on either side of the starting point of the integration contour. Only the second-order pole in such a collision lead to a divergence. These divergences read:
\begin{align}
\sum_{M=0}^\infty & (-)^M(-\log \epsilon) \int_{2\pi>\sigma_1>...>\sigma_M>0} d\sigma_1...d\sigma_M A^{B^{(1)}}(\sigma_1)...A^{B^{(M)}}(\sigma_M)\cr
& \qquad \times \sum_{m,n}(F_m F_n C_{mn} + \bar F_m \bar F_n C_{\bar m \bar n}) 
\kappa^{C_n D_m} t_{D_m} t_{B^{(1)}}...t_{B^{(M)}} t_{C_n} \prod_{i=1}^M(-)^{C_n B^{(i)}}
\end{align}
So the first-order divergences in the monodromy matrix add up to:
\begin{align}\label{divMonodromy} & \log \epsilon \frac{y^4 + y^{-4}}{2} 
\sum_{M=0}^{\infty} (-)^M \int_{2\pi>\sigma_1>...>\sigma_M>0} d\sigma_1...d\sigma_M A^{B^{(1)}}(\sigma_1)...A^{B^{(M)}}(\sigma_M)\cr
& \qquad \times ( -(\kappa^{C_3 D_1} t_{D_1} t_{C_3} + \kappa^{C_2 D_2} t_{D_2} t_{C_2} + \kappa^{C_1 D_3} t_{D_3} t_{C_1}) t_{B^{(1)}}...t_{B^{(M)}} \cr
& \qquad \qquad - t_{B^{(1)}}...t_{B^{(M)}} (\kappa^{C_3 D_1} t_{D_1} t_{C_3} + \kappa^{C_2 D_2} t_{D_2} t_{C_2} + \kappa^{C_1 D_3} t_{D_3} t_{C_1}) \cr
& \qquad \qquad + 2\kappa^{C_3 D_1} t_{D_1}t_{B^{(1)}}...t_{B^{(M)}} t_{C_3}\prod_{i=1}^M(-)^{D_1 B^{(i)}}
 + 2\kappa^{C_2 D_2} t_{D_2}t_{B^{(1)}}...t_{B^{(M)}} t_{C_2}\prod_{i=1}^M(-)^{D_2 B^{(i)}}\cr
& \qquad \qquad 
 + 2\kappa^{C_1 D_3} t_{D_3}t_{B^{(1)}}...t_{B^{(M)}} t_{C_1}\prod_{i=1}^M(-)^{D_3 B^{(i)}}
 )
\end{align}
Taking the supertrace, we see that the transfer matrix is free of divergences at first order.


\section{Fusion at second order: computations}\label{comFusion}

In this appendix we give some details concerning the computation of the fusion of line operators at second order. 
In particular we describe the computation that leads to \eqref{FusionTT1} and \eqref{FusionTT2}.
These terms are produced by triple collisions of connections.
A triple collision means that we take one OPE between two connections, and then take the OPE of the resulting currents with a third connection.

\paragraph{Treatment of the OPEs.}
As explain in section \ref{recipeFusion}, one needs to disentangle two contributions from the OPEs.
On one hand there is the contribution that gives a quantum correction associated with fusion.
On the other hand there is the contribution that is interpreted as a regularized OPE in the double line operator resulting from the process of fusion.
In order to isolate the interesting part associated with fusion, we subtract the principal value from the singularities. We obtain%
\footnote{The regularized delta-function in the third line of \eqref{regPoles} is not exactly the same one as in the first two lines. However to keep the formulas simple we will adopt the same notations for both regularizations of the delta-function.}%
:
\begin{align}\label{regPoles}
&\frac{1}{(\sigma \pm i \epsilon - \sigma')^2} -P.V. \frac{1}{(\sigma-\sigma')^2}  = \pm i \pi \delta'_\epsilon(\sigma-\sigma') \cr
&\frac{1}{\sigma \pm i \epsilon - \sigma'} - P.V. \frac{1}{\sigma-\sigma'} = \mp i \pi \delta_\epsilon(\sigma-\sigma')  \cr
&\frac{\sigma \mp i \epsilon - \sigma'}{(\sigma \pm i \epsilon - \sigma')^2} - P.V. \frac{1}{\sigma-\sigma'} = \mp i \pi \delta_\epsilon(\sigma-\sigma')  
\end{align}

\paragraph{Computation of the individual terms.}
Let us now face the computation of the individual quantum corrections that add up to \eqref{FusionTT1} and \eqref{FusionTT2}.
In the first step of the computation of a triple collision we perform an OPE between two connections sitting on different contours.
We obtain intermediate currents that we evaluate on one of the two contours\footnote{We can also choose to evaluate these intermediary currents in between the two contours. This would not change equations \eqref{KtildeLim} and \eqref{tttildeLim}.}.
 The contour on which these intermediate currents are evaluated matters for the second step of the computation. 
The relevant OPE for the first step of the computation is thus:
\begin{align}\label{firstOPE} (1&-P.V.) A_R(y;\sigma+i\epsilon)A_{R'}(y';\sigma') \supset \pi i R^{-2} \delta_\epsilon(\sigma-\sigma')\sum_{m,n=0}^3 \sum_{p} {f_{C_p}}^{b_n A_m} t_{A_m}^R t_{B_n}^{R'} \cr
& \times (D_{m n}^{p} K_p^{C_p}(\sigma+i\epsilon) + {D'}_{m n}^{p} K_p^{C_p}(\sigma') + 
D_{m n}^{\bar p} \bar K_p^{C_p}(\sigma+i\epsilon) + {D'}_{m n}^{\bar p} \bar K_p^{C_p}(\sigma') )
\end{align}
where the index $p$ can take the values $\{0,1,2,3,g\}$. In the following when the range of the sum for some index is not specified, it is understood that the sum runs over the set $\{0,1,2,3,g\}$.
The coefficients $D_{**}^{*}$ depends on the precise location where the currents are evaluated in the simple poles of the current algebra \eqref{Kalgebra}. From the coefficient given in section \ref{subKalgebra} we can only deduce the sums $ D_{m n}^{p} +  {D'}_{m n}^{p} $ and $ D_{m n}^{\bar p} +  {D'}_{m n}^{\bar p} $.
It turns out that we don't need more information about the coefficients $D_{**}^{*}$ for the purpose of this article (see equation \eqref{DsLim}).

Actually what we want to compute here is not exactly the contribution of the triple collisions to the fusion of line operators. 
It is rather the part of this contribution that has not been taken into account by the first part of the computation described in section \ref{subFusion2nd} and depicted in figure \ref{fusO2step1}. Removing the piece of the triple collisions already taken into account amounts to perform the following replacement in the first OPE \eqref{firstOPE}:
\begin{align}\label{defDtilde}
&D_{mn}^p \to \tilde D_{mn}^p = D_{mn}^p - \frac{1}{2}F_p(y)(s_{4-n,n}+r_{4-n,n}) \cr
&D_{mn}^{\bar p} \to \tilde D_{mn}^{\bar p} = D_{mn}^{\bar p} - \frac{1}{2}\bar F_p(y)(s_{4-n,n}+r_{4-n,n}) \cr
&{D'}_{mn}^p \to \tilde {D'}_{mn}^p = {D'}_{mn}^p - \frac{1}{2}F_p(y')(s_{m,4-m}-r_{m,4-m}) \cr
&{D'}_{mn}^{\bar p} \to \tilde {D'}_{mn}^{\bar p} = {D'}_{mn}^{\bar p} - \frac{1}{2}\bar F_p(y')(s_{m,4-m}-r_{m,4-m})
\end{align}

\begin{figure}
\centering
\includegraphics[scale=0.55]{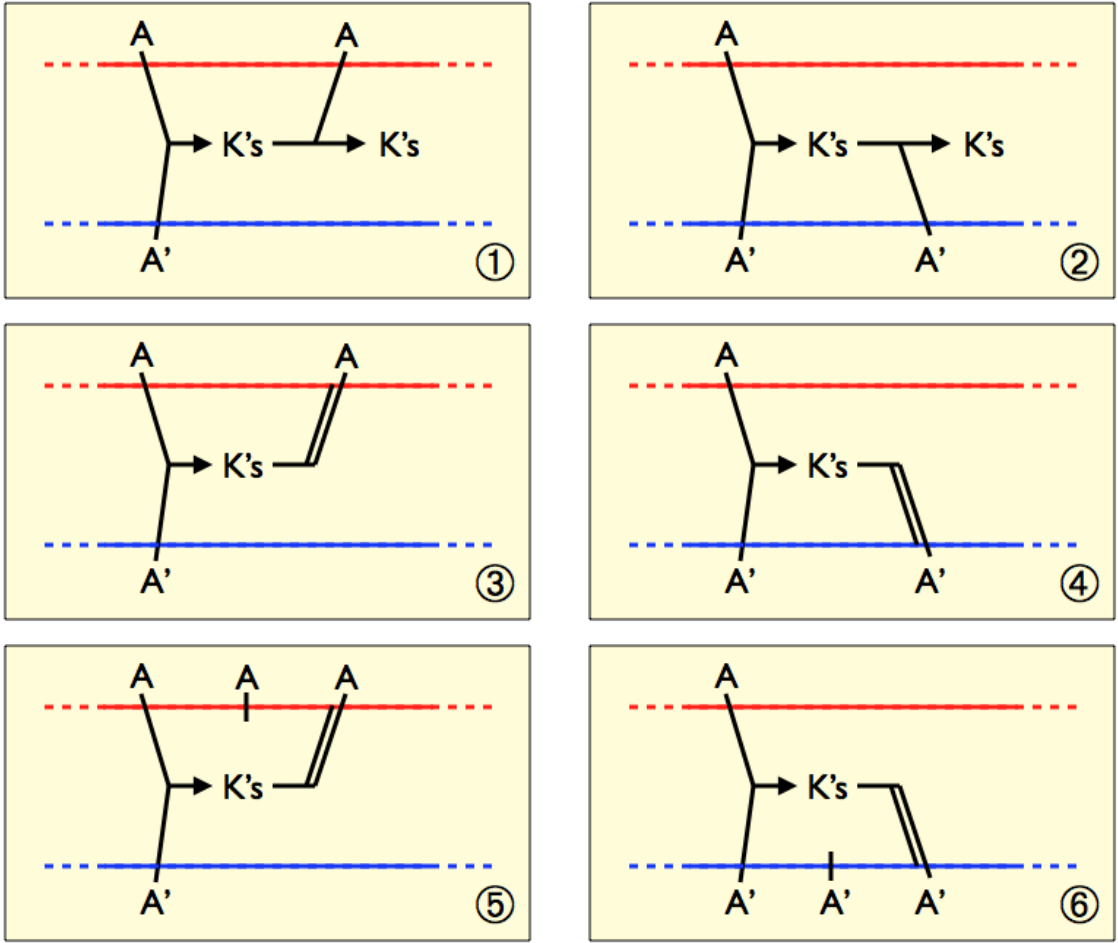}
\caption{The different types of triple collisions that contribute to the fusion at second order. The first OPE produces currents denoted $K's$. In \ding{172} and \ding{173} the contribution from the first-order poles in the second OPE is singled out. In the four other cases the contribution from the second-order poles in the second OPE is considered.\label{allTripleCollisions}}
\end{figure}

Now we can compute the individual terms that add up to \eqref{FusionTT1} and \eqref{FusionTT2}.
These different terms are schematically depicted in figure \ref{allTripleCollisions}.
We compute separately the contribution of the first- and second-order poles in the second OPE (obviously only the first order poles have to be taken into account in the first OPE).
Let us begin with the first-order poles in the second OPE. The triple collision involves three connections, two of which are integrated on the same contour. There are two different cases.
For a triple collision involving two neighboring connections on the first contour (case \ding{172} in figure \ref{allTripleCollisions}), we obtain:
\begin{align}\label{RRR'1}
 (i\pi R^{-2})^2 & \delta_{\epsilon}(\sigma_1-\sigma')\delta_{\epsilon}(\sigma_2-\sigma') \sum_{m,n=0}^3\sum_{p,q,r}
{f_{C_p}}^{B_n A_m} {f_{E_r}}^{C_p D_q} \{ t_{D_q}^R , t_{A_m}^R] t_{B_n}^{R'} \cr
& \times (K_r^{E_r}({\tilde{D'}}_{mn}^p F_q C_{pq}^r - {\tilde{D'}}_{mn}^p \bar F_q C_{p\bar q}^r - {\tilde{D'}}_{mn}^{\bar p} F_q C_{\bar pq}^r - {\tilde{D'}}_{mn}^{\bar p} \bar F_q C_{\bar p \bar q}^r ) \cr
& \quad + \bar K_r^{E_r}({\tilde{D'}}_{mn}^p F_q C_{pq}^{\bar r} + {\tilde{D'}}_{mn}^p \bar F_q C_{p\bar q}^{\bar r} + {\tilde{D'}}_{mn}^{\bar p} F_q C_{\bar pq}^{\bar r} - {\tilde{D'}}_{mn}^{\bar p} \bar F_q C_{\bar p \bar q}^{\bar r} ))
\end{align}
where we wrote $F_q$ as a shorthand for $F_q(y)$. Similarly we will write $F'_q$ for $F_q(y')$.
For a triple collision involving two neighboring connections on the second contour (case \ding{173} in figure \ref{allTripleCollisions}), we obtain:
\begin{align}\label{RR'R'1}
 (i\pi R^{-2})^2 & \delta_{\epsilon}(\sigma'_1-\sigma)\delta_{\epsilon}(\sigma'_2-\sigma) \sum_{m,n=0}^3\sum_{p,q,r}
{f_{C_p}}^{B_n A_m} {f_{E_r}}^{ D_q C_p}  t_{A_m}^R \{ t_{B_n}^{R'}, t_{D_q}^{R'}] \cr
& \times (K_r^{E_r}(-{\tilde{D}}_{mn}^p F'_q C_{pq}^r + {\tilde{D}}_{mn}^p \bar F'_q C_{p\bar q}^r + {\tilde{D}}_{mn}^{\bar p} F'_q C_{\bar pq}^r + {\tilde{D}}_{mn}^{\bar p} \bar F'_q C_{\bar p \bar q}^r ) \cr
& \quad + \bar K_r^{E_r}(-{\tilde{D}}_{mn}^p F'_q C_{pq}^{\bar r} - {\tilde{D}}_{mn}^p \bar F'_q C_{p\bar q}^{\bar r} - {\tilde{D}}_{mn}^{\bar p} F'_q C_{\bar pq}^{\bar r} + {\tilde{D}}_{mn}^{\bar p} \bar F'_q C_{\bar p \bar q}^{\bar r} ))
\end{align}
Next we consider the second-order poles in the second OPE. There are now four different cases.
For a triple collision involving two neighboring connections on the first contour (case \ding{174} in figure \ref{allTripleCollisions}), we obtain:
\begin{align}\label{RRR'2}
& (i\pi R^{-2})^2  \delta'_{\epsilon}(\sigma_1-\sigma')\delta_{\epsilon}(\sigma_2-\sigma') \sum_{m,n=0}^3\sum_{p,q}
{f}^{B_n A_m D_q}   t_{D_q}^R t_{A_m}^R  t_{B_n}^{R'} (-{\tilde{D'}}_{mn}^p F_q C_{pq} + {\tilde{D'}}_{mn}^{\bar p} \bar F_q C_{\bar p\bar q}) \cr
& +  (i\pi R^{-2})^2  \delta_{\epsilon}(\sigma_1-\sigma')\delta'_{\epsilon}(\sigma_2-\sigma') \sum_{m,n=0}^3\sum_{p,q}
{f}^{B_n A_m D_q}  (-)^{A_m D_q} t_{A_m}^R  t_{D_q}^R  t_{B_n}^{R'} (-{\tilde{D'}}_{mn}^p F_q C_{pq} + {\tilde{D'}}_{mn}^{\bar p} \bar F_q C_{\bar p\bar q})
\end{align}
and for a triple collision involving two neighboring connections on the second contour (case \ding{175} in figure \ref{allTripleCollisions}), we obtain:
\begin{align}\label{RR'R'2}
& -(i\pi R^{-2})^2  \delta_{\epsilon}(\sigma'_1-\sigma)\delta'_{\epsilon}(\sigma'_2-\sigma)  \sum_{m,n=0}^3\sum_{p,q}
{f}^{ D_q B_n A_m}  t_{A_m}^R  t_{B_n}^{R'}  t_{D_q}^{R'} (-{\tilde{D}}_{mn}^p F'_q C_{pq} + {\tilde{D}}_{mn}^{\bar p} \bar F'_q C_{\bar p\bar q}) \cr
& -  (i\pi R^{-2})^2  \delta'_{\epsilon}(\sigma'_1-\sigma)\delta_{\epsilon}(\sigma'_2-\sigma)  \sum_{m,n=0}^3\sum_{p,q}
{f}^{ D_q B_n A_m}  t_{A_m}^R  (-)^{D_q B_n} t_{D_q}^{R'} t_{B_n}^{R'} (-{\tilde{D}}_{mn}^p F'_q C_{pq} + {\tilde{D}}_{mn}^{\bar p} \bar F'_q C_{\bar p\bar q})
\end{align}
We also obtain a non-zero contribution if the two connections that are on the same contour are separated by a third one. 
When this happens on the first contour (case \ding{176} in figure \ref{allTripleCollisions}), we obtain:
\begin{align}\label{RARR'}
 (i\pi R^{-2})^2 & \delta'_{\epsilon}(\sigma_1-\sigma')\delta_{\epsilon}(\sigma_2-\sigma') \frac{1}{2} \sum_{m,n=0}^3\sum_{p,q}
\{ A_R(\sigma), {f}^{B_n A_m D_q}  [ t_{D_q}^R , t_{A_m}^R \} t_{B_n}^{R'} \} \cr
& \quad \times (-{\tilde{D'}}_{mn}^p F_q C_{pq} + {\tilde{D'}}_{mn}^{\bar p} \bar F_q C_{\bar p\bar q})\cr
+ (i\pi R^{-2})^2 & \delta'_{\epsilon}(\sigma_1-\sigma')\delta_{\epsilon}(\sigma_2-\sigma') \frac{1}{2} \sum_{m,n,q=0}^3 \sum_{p,r,s}
{f_{C_p}}^{B_n A_m} {f_{E_r}}^{C_p D_q} \{ t_{D_q}^R , t_{A_m}^R ] t_{B_n}^{R'}  \cr
& \quad \times ({\tilde{D'}}_{mn}^s F_p C_{sp} - {\tilde{D'}}_{mn}^{\bar s} \bar F_p C_{\bar s\bar p})
(F_r K_r^{E_r} + \bar F_r \bar K_r^{E_r})\cr
+ (i\pi R^{-2})^2 & \delta'_{\epsilon}(\sigma_1-\sigma')\delta_{\epsilon}(\sigma_2-\sigma') \frac{1}{2} \sum_{n,p,q=0}^3\sum_{m,r,s}
{f_{C_p}}^{B_n A_m} {f_{E_r}}^{C_p D_q} \{ t_{D_q}^R , t_{A_m}^R ] t_{B_n}^{R'}  \cr
& \quad \times ({\tilde{D'}}_{pn}^s F_m C_{sm}  - {\tilde{D'}}_{pn}^{\bar s} \bar F_m C_{\bar s\bar m})
(F_r K_r^{E_r} + \bar F_r \bar K_r^{E_r})
\end{align}
Finally for two connections separated by a third one in the second contour (case \ding{177} in figure  \ref{allTripleCollisions}), we obtain:
\begin{align}\label{RR'A'R'}
 (i\pi R^{-2})^2 & \delta'_{\epsilon}(\sigma'_1-\sigma)\delta_{\epsilon}(\sigma'_2-\sigma) \frac{1}{2} \sum_{m,n=0}^3\sum_{p,q}
\{ A_{R'}(\sigma'), {f}^{D_q B_n A_m}  t_{A_m}^R [ t_{B_n}^{R'},t_{D_q}^{R'}\} \} \cr
& \quad \times (-{\tilde{D}}_{mn}^p F'_q C_{pq} + {\tilde{D}}_{mn}^{\bar p} \bar F'_q C_{\bar p\bar q})\cr
+ (i\pi R^{-2})^2 & \delta'_{\epsilon}(\sigma'_1-\sigma)\delta_{\epsilon}(\sigma'_2-\sigma) \frac{1}{2} \sum_{m,n,q=0}^3\sum_{p,r,s}
{f_{C_p}}^{B_n A_m} {f_{E_r}}^{D_q C_p} t_{A_m}^R \{ t_{B_n}^{R'},t_{D_q}^{R'}]  \cr
& \quad \times ({\tilde{D}}_{mn}^s F'_p C_{sp}  - {\tilde{D}}_{mn}^{\bar s} \bar F'_p C_{\bar s\bar p} )
(F'_r K_r^{E_r} + \bar F'_r \bar K_r^{E_r})\cr
+ (i\pi R^{-2})^2 & \delta'_{\epsilon}(\sigma'_1-\sigma)\delta_{\epsilon}(\sigma'_2-\sigma) \frac{1}{2} \sum_{n,p,q=0}^3 \sum_{m,r,s}
{f_{C_p}}^{B_n A_m} {f_{E_r}}^{D_q C_p} t_{A_m}^R \{ t_{B_n}^{R'},t_{D_q}^{R'}]  \cr
& \quad \times ({\tilde{D}}_{pn}^s F'_m C_{sm} - {\tilde{D}}_{pn}^{\bar s} \bar F'_m C_{\bar s\bar m})
(F'_r K_r^{E_r} + \bar F'_r \bar K_r^{E_r})\
\end{align}

\paragraph{Performing the integration.}
Next we have to perform the integration over the free coordinates in the previous results.
The integrals over the regularized delta functions provide a well-defined answer. This is an advantage of the OPE formalism with respect to the Hamiltonian formalism.
The integrals needed are given below.
The results are given in the limit $\epsilon \to 0$.
\begin{align}
\int_{b>\sigma_1>\sigma_2>a}d\sigma_1 d\sigma_2 \delta_\epsilon(\sigma_1-\sigma')\delta_\epsilon(\sigma_2-\sigma') = \frac{1}{2} \chi(\sigma';a,b)
\end{align}
\begin{align}
\int_{b>\sigma_1>\sigma>\sigma_2>a}d\sigma_1 d\sigma_2 \int_c^d d\sigma' \delta'_\epsilon(\sigma_1-\sigma')\delta_\epsilon(\sigma_2-\sigma') = -\frac{1}{2} \chi(\sigma;c,d)
\end{align}
\begin{align}
& \int_{b>\sigma_1>\sigma_2>a}d\sigma_1 d\sigma_2 \int_c^d d\sigma' \delta'_\epsilon(\sigma_1-\sigma')\delta_\epsilon(\sigma_2-\sigma') \cr
&\qquad = \frac{1}{2} \chi(b;c,d) - \int_{a}^b d\sigma_2 \int_c^d d\sigma'  \delta^2_\epsilon(\sigma_2-\sigma')
\end{align}
It is sometimes convenient to write $\chi(b;c,d)$ as $\frac{1}{2}( \chi(b;c,d) + \chi(a;c,d) + \chi(c;a,b) - \chi(d;a,b) )$.
Notice that the integral over the squared regularized delta function is divergent in the limit $\epsilon \to 0$:
\be \int_{a}^b d\sigma_2 \int_c^d d\sigma'  \delta^2_\epsilon(\sigma_2-\sigma')
= \frac{1}{2\pi \epsilon}|[a,b]\cap[c,d]| \ee
where we denoted by $|[a,b]\cap[c,d]|$ the length of the overlap of the intervals $[a,b]$ and $[c,d]$.
Similarly we have:
\begin{align}
& \int_{b>\sigma_1>\sigma_2>a}d\sigma_1 d\sigma_2 \int_c^d d\sigma' \delta_\epsilon(\sigma_1-\sigma')\delta'_\epsilon(\sigma_2-\sigma') = -\frac{1}{2} \chi(a;c,d) + \int_{a}^b d\sigma_2 \int_c^d d\sigma'  \delta^2_\epsilon(\sigma_2-\sigma')\end{align}
where we can also replace $\chi(a;c,d)$ by $\frac{1}{2}( \chi(b;c,d) + \chi(a;c,d) - \chi(c;a,b) + \chi(d;a,b) ) $.

\paragraph{Summing the terms.}
Finally we can sum the various contributions from triple collisions.
The terms of the form \eqref{RRR'1} combined with the second and third terms of  \eqref{RARR'} lead to:
\begin{align}\label{fusionttt'}
\sum_{M=0}^\infty& (-)^{M+M'+3} \frac{1}{2} \sum_{i=0}^M \sum_{i'=0}^{M'} \int_{[a,b]\cap[c,d]}d\sigma 
\left \lfloor \int_\sigma^b A \right\rceil^i \left \lfloor \int_\sigma^d A' \right\rceil^{i'} \cr
& \times  (i\pi R^{-2})^2  \tilde{\sum_{m,n,p,q,r}}
{f_{C_p}}^{B_n A_m} {f_{E_r}}^{C_p D_q} \{ t_{D_q}^R , t_{A_m}^R] t_{B_n}^{R'} \cr
&\quad \times (K_r^{E_r}({\tilde{D'}}_{mn}^p F_q C_{pq}^r - {\tilde{D'}}_{mn}^p \bar F_q C_{p\bar q}^r - {\tilde{D'}}_{mn}^{\bar p} F_q C_{\bar pq}^r - {\tilde{D'}}_{mn}^{\bar p} \bar F_q C_{\bar p \bar q}^r \cr
&\quad \quad\quad+ \frac{1}{2} F_r \sum_s ({\tilde{D'}}_{mn}^s F_p C_{sp} + {\tilde{D'}}_{pn}^s F_m C_{sm} - {\tilde{D'}}_{mn}^{\bar s} \bar F_p C_{\bar s\bar p} - {\tilde{D'}}_{pn}^{\bar s} \bar F_m C_{\bar s\bar m})) \cr
& \quad\quad + \bar K_r^{E_r}({\tilde{D'}}_{mn}^p F_q C_{pq}^{\bar r} + {\tilde{D'}}_{mn}^p \bar F_q C_{p\bar q}^{\bar r} + {\tilde{D'}}_{mn}^{\bar p} F_q C_{\bar pq}^{\bar r} - {\tilde{D'}}_{mn}^{\bar p} \bar F_q C_{\bar p \bar q}^{\bar r} \cr
& \quad\quad\quad+ \frac{1}{2} \bar F_r \sum_s ({\tilde{D'}}_{mn}^s F_p C_{sp} + {\tilde{D'}}_{pn}^s F_m C_{sm} - {\tilde{D'}}_{mn}^{\bar s} \bar F_p C_{\bar s\bar p} - {\tilde{D'}}_{pn}^{\bar s} \bar F_m C_{\bar s\bar m}))) \cr
& \times \left \lfloor \int_a^\sigma A \right\rceil^{M-i} \left \lfloor \int_c^\sigma A' \right\rceil^{M'-i'} 
\end{align}
In order to shorten the previous expression we introduce the symbol $\tilde{\sum}$ with the following meaning: for each term in the expression, the lower indices of a coefficient $D$ have to be summed over the values $\{0,1,2,3\}$, while all other indices have to be summed over the values $\{0,1,2,3,g\}$. 
Similarly the terms of the form \eqref{RR'R'1} combined with the second and third terms of  \eqref{RR'A'R'} lead to:
\begin{align}\label{fusiontt't'}
\sum_{M=0}^\infty& (-)^{M+M'+3} \frac{1}{2} \sum_{i=0}^M \sum_{i'=0}^{M'} \int_{[a,b]\cap[c,d]}d\sigma 
\left \lfloor \int_\sigma^b A \right\rceil^i \left \lfloor \int_\sigma^d A' \right\rceil^{i'} \cr
& \times  (i\pi R^{-2})^2  \tilde{\sum_{m,n,p,q,r}}
{f_{C_p}}^{B_n A_m} {f_{E_r}}^{ D_q C_p}  t_{A_m}^R \{ t_{B_n}^{R'}, t_{D_q}^{R'}]  \cr
&\quad \times (K_r^{E_r}(-{\tilde{D}}_{mn}^p F'_q C_{pq}^r + {\tilde{D}}_{mn}^p \bar F'_q C_{p\bar q}^r + {\tilde{D}}_{mn}^{\bar p} F'_q C_{\bar pq}^r + {\tilde{D}}_{mn}^{\bar p} \bar F'_q C_{\bar p \bar q}^r \cr
&\quad \quad\quad- \frac{1}{2} F'_r \sum_s ({\tilde{D}}_{mn}^s F'_p C_{sp} + {\tilde{D}}_{pn}^s F'_m C_{sm} - {\tilde{D}}_{mn}^{\bar s} \bar F'_p C_{\bar s\bar p} - {\tilde{D}}_{pn}^{\bar s} \bar F'_m C_{\bar s\bar m})) \cr
& \quad\quad + \bar K_r^{E_r}(-{\tilde{D}}_{mn}^p F'_q C_{pq}^{\bar r} - {\tilde{D}}_{mn}^p \bar F'_q C_{p\bar q}^{\bar r} - {\tilde{D}}_{mn}^{\bar p} F'_q C_{\bar pq}^{\bar r} + {\tilde{D}}_{mn}^{\bar p} \bar F'_q C_{\bar p \bar q}^{\bar r} ) \cr
& \quad\quad\quad- \frac{1}{2} \bar F'_r \sum_s ({\tilde{D}}_{mn}^s F'_p C_{sp} + {\tilde{D}}_{pn}^s F'_m C_{sm} - {\tilde{D}}_{mn}^{\bar s} \bar F'_p C_{\bar s\bar p} - {\tilde{D}}_{pn}^{\bar s} \bar F'_m C_{\bar s\bar m}))) \cr
& \times \left \lfloor \int_a^\sigma A \right\rceil^{M-i} \left \lfloor \int_c^\sigma A' \right\rceil^{M'-i'} 
\end{align}
The sum of \eqref{fusionttt'} and \eqref{fusiontt't'} leads to \eqref{FusionTT1} where the operator $\tilde K$ is given by:
\begin{align}\label{Ktilde}
\tilde K = \frac{\pi^2}{2} &\tilde{\sum_{m,n,p,q,r}}
{f_{C_p}}^{B_n A_m} {f_{E_r}}^{C_p D_q} \{ t_{D_q}^R , t_{A_m}^R] t_{B_n}^{R'} \cr
&\quad \times (K_r^{E_r}({\tilde{D'}}_{mn}^p F_q C_{pq}^r - {\tilde{D'}}_{mn}^p \bar F_q C_{p\bar q}^r - {\tilde{D'}}_{mn}^{\bar p} F_q C_{\bar pq}^r - {\tilde{D'}}_{mn}^{\bar p} \bar F_q C_{\bar p \bar q}^r \cr
&\quad \quad\quad+ \frac{1}{2} F_r  \sum_s ({\tilde{D'}}_{mn}^s F_p C_{sp} + {\tilde{D'}}_{pn}^s F_m C_{sm} - {\tilde{D'}}_{mn}^{\bar s} \bar F_p C_{\bar s\bar p} - {\tilde{D'}}_{pn}^{\bar s} \bar F_m C_{\bar s\bar m})) \cr
& \quad\quad + \bar K_r^{E_r}({\tilde{D'}}_{mn}^p F_q C_{pq}^{\bar r} + {\tilde{D'}}_{mn}^p \bar F_q C_{p\bar q}^{\bar r} + {\tilde{D'}}_{mn}^{\bar p} F_q C_{\bar pq}^{\bar r} - {\tilde{D'}}_{mn}^{\bar p} \bar F_q C_{\bar p \bar q}^{\bar r} \cr
& \quad\quad\quad+ \frac{1}{2} \bar F_r  \sum_s ({\tilde{D'}}_{mn}^s F_p C_{sp} + {\tilde{D'}}_{pn}^s F_m C_{sm} - {\tilde{D'}}_{mn}^{\bar s} \bar F_p C_{\bar s\bar p} - {\tilde{D'}}_{pn}^{\bar s} \bar F_m C_{\bar s\bar m})))\cr
&+{f_{C_p}}^{B_n A_m} {f_{E_r}}^{ D_q C_p}  t_{A_m}^R \{ t_{B_n}^{R'}, t_{D_q}^{R'}]  \cr
&\quad \times (K_r^{E_r}(-{\tilde{D}}_{mn}^p F'_q C_{pq}^r + {\tilde{D}}_{mn}^p \bar F'_q C_{p\bar q}^r + {\tilde{D}}_{mn}^{\bar p} F'_q C_{\bar pq}^r + {\tilde{D}}_{mn}^{\bar p} \bar F'_q C_{\bar p \bar q}^r \cr
&\quad \quad\quad- \frac{1}{2} F'_r  \sum_s ({\tilde{D}}_{mn}^s F'_p C_{sp} + {\tilde{D}}_{pn}^s F'_m C_{sm} - {\tilde{D}}_{mn}^{\bar s} \bar F'_p C_{\bar s\bar p} - {\tilde{D}}_{pn}^{\bar s} \bar F'_m C_{\bar s\bar m})) \cr
& \quad\quad + \bar K_r^{E_r}(-{\tilde{D}}_{mn}^p F'_q C_{pq}^{\bar r} - {\tilde{D}}_{mn}^p \bar F'_q C_{p\bar q}^{\bar r} - {\tilde{D}}_{mn}^{\bar p} F'_q C_{\bar pq}^{\bar r} + {\tilde{D}}_{mn}^{\bar p} \bar F'_q C_{\bar p \bar q}^{\bar r} ) \cr
& \quad\quad\quad- \frac{1}{2} \bar F'_r  \sum_s ({\tilde{D}}_{mn}^s F'_p C_{sp} + {\tilde{D}}_{pn}^s F'_m C_{sm} - {\tilde{D}}_{mn}^{\bar s} \bar F'_p C_{\bar s\bar p} - {\tilde{D}}_{pn}^{\bar s} \bar F'_m C_{\bar s\bar m})))
\end{align}
The combination of the terms \eqref{RRR'2} and \eqref{RR'R'2} together with the first terms of \eqref{RARR'} and \eqref{RR'A'R'} simplifies to:
\begin{align}\label{fusiontt'}
\sum_{M=0}^\infty& (-)^{M+M'+3} \frac{1}{2} \sum_{i=0}^M \sum_{i'=0}^{M'} \int_{a}^b d\sigma \int_{c}^d d\sigma' 
\left \lfloor \int_{\sigma}^b A \right\rceil^i \left \lfloor \int_{\sigma'}^d A' \right\rceil^{i'} \cr
& \times  (i\pi R^{-2})^2  \delta_\epsilon^2(\sigma-\sigma') \tilde{\sum_{m,n,p,q,r}} f^{B_n A_m D_q} {f_{D_q A_m}}^{E_r} t_{E_r}^R t_{B_n}^{R'}\cr
& \quad (-{\tilde{D'}}_{mn}^p F_q C_{pq} + {\tilde{D'}}_{mn}^{\bar p} \bar F_q C_{\bar p\bar q}
-{\tilde{D}}_{rm}^p F'_q C_{pq} + {\tilde{D}}_{rm}^{\bar p} \bar F'_q C_{\bar p\bar q})\cr
& \times \left \lfloor \int_a^{\sigma} A \right\rceil^{M-i} \left \lfloor \int_c^{\sigma'} A' \right\rceil^{M'-i'} 
\end{align}
This can be rewritten as \eqref{FusionTT2} where the matrix $\tilde{tt}$ is given by:
\begin{align}\label{tttilde}
\tilde{tt}=&\frac{\pi^2}{2} \tilde{\sum_{m,n,p,q,r}} f^{B_n A_m D_q} {f_{D_q A_m}}^{E_r} t_{E_r}^R t_{B_n}^{R'}
\cr& \quad 
\times (-{\tilde{D'}}_{mn}^p F_q C_{pq} + {\tilde{D'}}_{mn}^{\bar p} \bar F_q C_{\bar p\bar q}
-{\tilde{D}}_{rm}^p F'_q C_{pq} + {\tilde{D}}_{rm}^{\bar p} \bar F'_q C_{\bar p\bar q})
\end{align}

\paragraph{Simplifications in the limit $y-y'=\mathcal{O}(R^{-2})$.}
For the purposes of this paper it is interesting to take the limit where the difference of spectral parameters is small. More precisely we assume that the difference $y-y'$ is of order $\mathcal{O}(R^{-2})$. In this limit one term dominates the previous result. This follows essentially from the observation that the $r$ matrix satisfies \eqref{rLim}. In the limit $y-y'=\mathcal{O}(R^{-2})$, the $r$ matrix is no longer of order $R^{-2}$ but rather of order $R^0$. Consequently the coefficients $\tilde{D}_{**}^*$ introduced in \eqref{defDtilde} behave like:
\begin{align}\label{DsLim}
& \tilde{D}_{mn}^p = -\frac{1}{4}F_p(y)\frac{y(y^2+y^{-2})^2}{y-y'} + \mathcal{O}(R^0) \cr
& \tilde{D}_{mn}^{\bar p} = -\frac{1}{4}\bar F_p(y)\frac{y(y^2+y^{-2})^2}{y-y'} + \mathcal{O}(R^0) \cr
& \tilde{D'}_{mn}^p = +\frac{1}{4}F_p(y)\frac{y(y^2+y^{-2})^2}{y-y'} + \mathcal{O}(R^0) \cr
& \tilde{D}_{mn}^{\bar p} = +\frac{1}{4}\bar F_p(y)\frac{y(y^2+y^{-2})^2}{y-y'} + \mathcal{O}(R^0) 
\end{align}
Using equation \eqref{eqsForRMatrix}, we deduce that \eqref{fusionttt'} simplifies to:
\begin{align}\label{fusionttt'Lim}
& (-)^{M+M'+3} \frac{1}{2} \sum_{i=0}^M \sum_{i'=0}^{M'} \int_{[a,b]\cap[c,d]}d\sigma 
\left \lfloor \int_\sigma^b A \right\rceil^i \left \lfloor \int_\sigma^d A' \right\rceil^{i'} \cr
& \times  (i\pi R^{-2})^2  \sum_{m,n,p,q=0}^3\sum_{r}
{f_{C_p}}^{B_n A_m} {f_{E_r}}^{C_p D_q} \{ t_{D_q}^R , t_{A_m}^R] t_{B_n}^{R'} \cr
&\quad \times \left( -\frac{1}{16}\frac{y^2(y^2+y^{-2})^4}{y-y'}\right) \left(\p_y F_r(y) K_r^{E_r} + \p_y \bar F_r(y) \bar K_r^{E_r}\right) \cr
& \times \left \lfloor \int_a^\sigma A \right\rceil^{M-i} \left \lfloor \int_c^\sigma A' \right\rceil^{M'-i'} +\mathcal{O}(R^{-4})
\end{align}
Remarkably the combination of currents that factors out is the derivative of the components of the flat connection \eqref{defAwithFs} with respect to the spectral parameter.
Similarly \eqref{fusiontt't'} simplifies to:
\begin{align}\label{fusiontt't'Lim}
& (-)^{M+M'+3} \frac{1}{2} \sum_{i=0}^M \sum_{i'=0}^{M'} \int_{[a,b]\cap[c,d]}d\sigma 
\left \lfloor \int_\sigma^b A \right\rceil^i \left \lfloor \int_\sigma^d A' \right\rceil^{i'} \cr
& \times  (i\pi R^{-2})^2  \sum_{m,n,p,q=0}^3\sum_{r}
{f_{C_p}}^{B_n A_m} {f_{E_r}}^{ D_q C_p}  t_{A_m}^R \{ t_{B_n}^{R'}, t_{D_q}^{R'}]  \cr
&\quad \times \left( -\frac{1}{16}\frac{y^2(y^2+y^{-2})^4}{y-y'}\right) \left(\p_y F_r(y) K_r^{E_r} + \p_y \bar F_r(y) \bar K_r^{E_r}\right) \cr
& \times \left \lfloor \int_a^\sigma A \right\rceil^{M-i} \left \lfloor \int_c^\sigma A' \right\rceil^{M'-i'} +\mathcal{O}(R^{-4})
\end{align}
So the operator $\tilde K$ becomes:
\begin{align}\label{KtildeLim}
 \tilde{K} &= - \frac{\pi^2}{32}\frac{y^2(y^2+y^{-2})^4}{y-y'}
   \sum_{m,n,p,q=0}^3\sum_{r} \p_y A^{E_r}(y)\cr
&  \times \left( {f_{C_p}}^{B_n A_m} {f_{E_r}}^{C_p D_q} \{ t_{D_q}^R , t_{A_m}^R] t_{B_n}^{R'}
+{f_{C_p}}^{B_n A_m} {f_{E_r}}^{ D_q C_p}  t_{A_m}^R \{ t_{B_n}^{R'}, t_{D_q}^{R'}] \right)
+\mathcal{O}(R^{0})
\end{align}
Eventually the term \eqref{fusiontt'} remains of order $R^{-4}$, since:
\be\label{tttildeLim} -{\tilde{D'}}_{mn}^p F_q C_{pq} + {\tilde{D'}}_{mn}^{\bar p} \bar F_q C_{\bar p\bar q}
-{\tilde{D}}_{rm}^p F'_q C_{pq} + {\tilde{D}}_{rm}^{\bar p} \bar F'_q C_{\bar p\bar q}
= 0 +\mathcal{O}(R^{-4})
\ee
Consequently the matrix $\tilde{tt}$ defined in \eqref{tttilde} remains of order one.
Actually it might be that the term \eqref{fusiontt'} and thus the matrix $\tilde{tt}$ cancel exactly. This could be decided by computing the coefficients of the derivatives of the currents in the current-current OPEs.



\end{appendix}

\end{document}